\documentclass[12pt]{article}

\usepackage{scicite}
\usepackage{graphicx}

\usepackage{times}

\newenvironment{sciabstract}{%
\begin{quote} \bf}
{\end{quote}}

\newcounter{lastnote}

\title{Antidepressant Use and Spatial Social Capital} 

\author
{Balázs Lengyel,$^{1,2\ast}$ Gergő Tóth,$^{1,3}$ Nicholas A. Christakis,$^{4}$ Anikó Bíró$^{5}$\\
\\
\normalsize{$^{1}$ANET Lab, Hungarian Research Network, 1097 Budapest, Hungary}\\
\normalsize{$^{2}$NETI Lab, Corvinus University of Budapest, 1093 Budapest, Hungary}\\
\normalsize{$^{3}$CERUM, Umeå University, 901 87 Umeå, Sweden}\\
\normalsize{$^{4}$Human Nature Lab, Yale University, 06520-8263 New Haven CT, USA}\\
\normalsize{$^{5}$Health and Population Research Group, Hungarian Research Network, 1097 Budapest, Hungary}\\
\\
\normalsize{$^\ast$To whom correspondence should be addressed; E-mail:  lengyel.balazs@krtk.hun-ren.hu.}
}

\date{}

\begin{document} 


\baselineskip24pt


\maketitle


\begin{sciabstract}
Social capital may help individuals maintain their mental health. Most empirical work based on small-scale surveys finds that bonding social capital and cohesive social networks are critical for mental well-being, while bridging social capital and diverse networks are considered less important. Here, we link data on antidepressant use of 277,344 small-town residents to a nation-wide online social network. The data enable us to examine how individuals' mental healthcare is related to the spatial characteristics of their social networks including their strong and weak ties. We find that, besides the cohesion of social networks around home, the diversity of connections to distant places is negatively correlated with the probability of antidepressant use. Spatial diversity of social networks is also associated with decreasing dosage in subsequent years. This relationship is independent from the local access to antidepressants and is more prevalent for young individuals. Structural features of spatial social networks are prospectively associated with depression treatment.
\end{sciabstract}

\clearpage

\section*{Main}

\noindent The theory of social capital \cite{putnam2000bowling, coleman1988social} has been used to help understand mental health inequalities \cite{mckenzie2002social, henderson2003social}. Relatedly, social support interventions have become part of health policy directed at mental disorders \cite{almedom2005social}. A rich literature suggests that social connections (the structural form of social capital) can reduce stress, anxiety, and depression \cite{rosenquist2011social, elmer2020students, santini2020social, smith2008, greenblatt1982social, kawachi2001social}. 
The central tenet is that mental balance can be primarily maintained with the help of emotional support gained from cohesive networks where individuals can access bonding social capital and ask help from their strong ties \cite{lin2017building}. 
Fewer investigators have argued for or explored the importance of bridging social capital \cite{fiori2006, salehi2019bonding}, that can be mobilized through weak ties and diverse networks, despite its pivotal role in providing, say, economic opportunities \cite{granovetter1973strength, eagle2010network, chetty2022social} that subsequently influence health outcomes \cite{thomson2022income}. A potential reason for the paucity of investigations of this kind is that previous empirical work has mostly used small-scale surveys to collect social connections and could not map the full horizon and extent of weak ties \cite{granovetter1973strength, dunbar1998social}. These data limitations have made the mental health outcomes of bonding versus bridging social capital difficult to compare. 

Here, we examine a nationwide data set on antidepressant use linked to the online social network of 277,344 small town residents; this provides us an informative scale for the social network analysis of mental health. We exploit the 
spatial dimension of connections which helps us disentangle bonding and bridging ties in the network. Our approach is based on the phenomenon that social capital falls with distance due to higher maintenance costs \cite{glaeser2002economic, borgatti2009network} and the empirical fact that the probability of links and network cohesion decreases with distance \cite{liben2005geographic, lambiotte2008geographical, lengyel2015geographies}, all suggesting that remote connections are likely weak. Furthermore, the spatial diversity of social connections correlates with individual economic wealth indicating that links across towns can be associated with bridging social capital \cite{eagle2010network}.
We quantify bonding social capital with a local cohesion (LC) indicator, that captures the tendency to participate in cohesive networks around subjects' homes, and we measure bridging social capital with a spatial diversity (SD) index that captures the capacity to connect diverse communities in distant places. 

We find that LC as well as SD predict the probability of antidepressant use with negative coefficients and confirm that both bonding and bridging social capital are related to mental health. However, the significance of local cohesion as a predictor of antidepressant use vanishes at high levels of spatial diversity suggesting that bridging can compensate for the lack of bonding social capital in the local environment. Furthermore, the dosage of antidepressants in subsequent years decreased more for patients who have spatially diverse networks. This latter relationship is stable after controlling for the local accessibility of antidepressants and is stronger for young individuals than for the elderly. Overall, the results suggest that bridging social capital and weak ties have a more important role in mental healthcare than previously thought.

\section*{Results}
\subsection*{Data}


\begin{figure*}[!ht]
\centering\includegraphics[width=\textwidth]{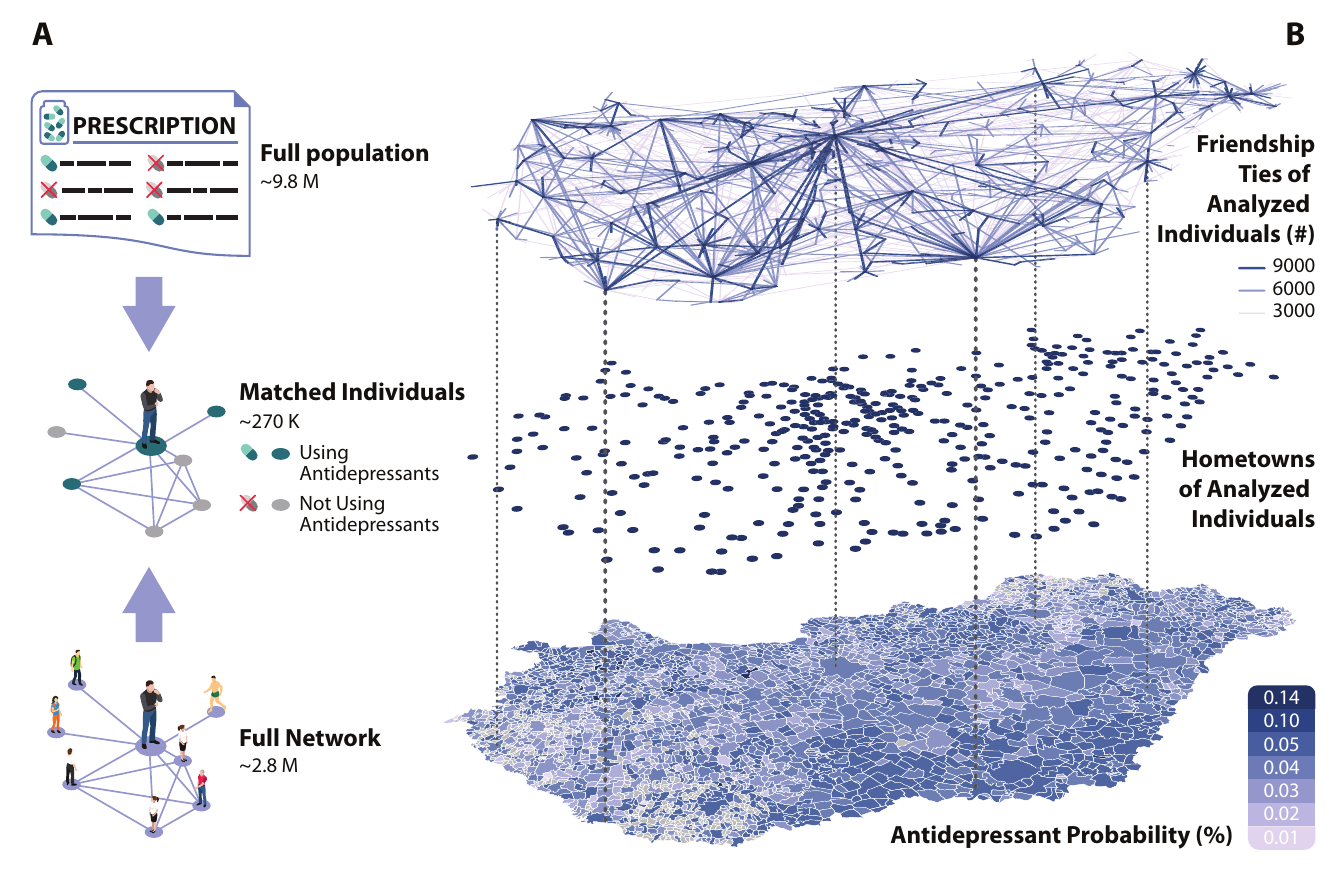}
\caption{The preparation and geographies of the data. \textbf{A.} A nation-wide dataset of antidepressant prescriptions is linked to the online social network of small-town residents in Hungary at the individual level. \textbf{B.} Hometowns of analyzed individuals and geographies of their social networks and local ratio of antidepressant users.}
\label{fig1}
\end{figure*}

We link antidepressant use to an online social network (OSN) on the individual level (Figure 1). Antidepressant use is retrieved from prescriptions that cover the entire population of Hungary in the 2011-2015 period and we also observe the dosage of antidepressants that the patients purchased. This prescription data does not capture undiagnosed mental problems or treatments without medication, but it does indicate depression, anxiety, or sleeping disorders \cite{pulkki2012, sinokki2009, gardarsdottir2007, oksanen2011} on a population scale. The OSN data comes from the iWiW social media portal that counted for 30\% of the country's total population in 2011. The iWiW data is a collection of friends, including their self-reported place of residency \cite{lengyel2015geographies}, and they have been previously used to measure the relationship of bonding and bridging social capital with corruption \cite{wachs2019social}, and income inequality \cite{toth2021inequality}. Our data is similar to other OSN sources that measure social capital at scale \cite{chetty2022social, bailey2018social}. Actual postings were rare on iWiW since it mainly functioned as a phone book and so was not very likely to induce anxiety like forms of social media that emerged later on \cite{steinfield2008social, shakya2017association, braghieri2022social}. 

We restrict the prescription data set to all residents of 409 Hungarian small towns with a population size between 5,000 and 20,000 which enables the probabilistic matching with the OSN data. We describe the ethical permissions and the matching  process in the Data and Methods section and provide further details on matching quality in Supporting Information 2. The resulting data set contains 277,344 individuals with zero or positive prescription drug use over the 2011-2015 period and network characteristics for the year 2011 (Figure 1A). These analyzed individuals live in towns both of high and low ratio of antidepressant users; and are distributed across the country (Figure 1B). The spatial structure of the analyzed social networks resemble the geographies of the entire iWiW population as it was documented before \cite{lengyel2015geographies}. A town-level analysis described in detail in Supporting Information 3 confirms that the sample towns do not significantly differ from the rest of the country in terms of the fraction of antidepressant users (two-sided Mann-Whitney U test, $p=0.971$).

\subsection*{Variables}
We define Spatial Social Capital as the geographical dimension of bonding and bridging social capital. The spatial structure of ego networks constructed from the full iWiW network is used to quantify bonding and bridging. That is, we include iWiW users living in all Hungarian towns to construct ego networks of analyzed individuals $i$. Detailed description and illustration of these variables can be found in Supporting Information 4.

The spatial dimension of bonding social capital is quantified by the Local Cohesion of social ties $LC_{i}$ as a function of network clustering. $LC_{i}$ measures the degree to which the $j$ and $k$ connections of individual $i$ residing in $i$'s hometown $h$ are linked to each other:
\begin{equation}
LC_{i} = \frac{2L_{jk}\in{h}} {n_{i}\in{h} ((n_{i}\in{h}) -1 )} / C_{i}^{ER}
\end{equation}
where $L_{jk}\in{h}$ is the number of links among $n_{i}\in{h}$ connections of $i$ ($j,k \in n_i$). To compare clustering coefficients of ego-networks of various sizes, we normalise the observed clustering of $i$ by the average clustering in $i$'s ego-network in ten simulated Erdős-Rényi ($ER$) random networks of $n_{i}$ nodes and $L_{jk}$ connections ($C_{i}^{ER}$), in which clustering is independent of degree \cite{neal2017small}.  Higher values of $LC_{i}$ imply that $i$ belongs to a strongly-knit local community compared to random networks of identical size, while low values imply that $i$ does not have a cohesive social network in her home town, taking into account the town's size (Figure 2A).

\begin{figure*}[!t]
\centering\includegraphics[width=\textwidth]{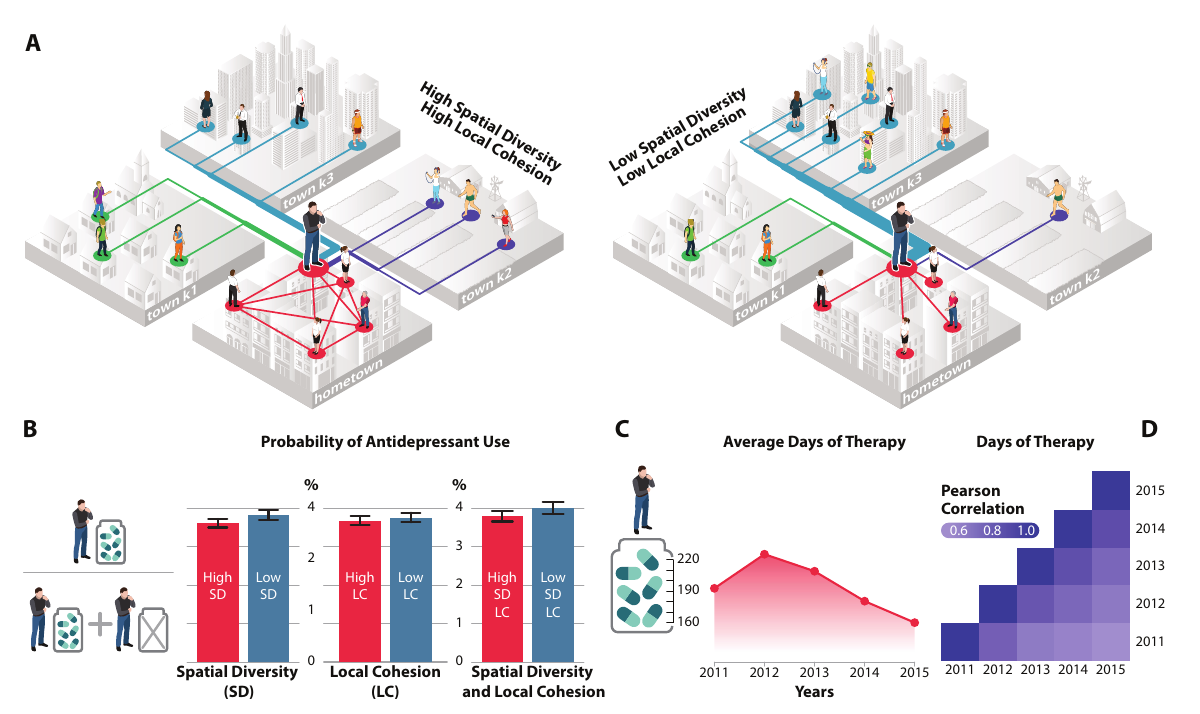}
\caption{Ego's Spatial Social Capital is quantified by the Local Cohesion of social network within Ego's town and by the Spatial Diversity of connections across towns.}
\label{fig}
\end{figure*}

Following \cite{eagle2010network}, we quantify bridging social capital by Spatial Diversity $SD_{i}$ in the ego network of individual $i$ using the Shannon entropy of connections across towns that we normalise by its theoretical maximum:
\begin{equation}
    SD_{i} = \frac{-\sum_{g=1}^{G}p_{ig}  \textrm{ln}(p_{ig})}{\textrm{ln}(G_i)}
\end{equation}
where $G_i$ is the number of different towns where connections of $i$ live ($h_i\in{G_i}$), $p_{ig}$ is the proportion of the total number of $i$ connections to town $g$.
$ p_{ig}= L_{ig} /  \sum{L_{i}}$
where $L_{ig}$ is the number of social ties of $i$ to town $g$, and the denominator is the total number of friendships of $i$. High Spatial Diversity implies that the individual splits their social connections more evenly among other towns, while low values imply that the individual's ties are concentrated in a relatively small number of towns (Figure 2A). 




Antidepressant use is captured by the binary variable $A_{i} \in(0,1)$ ($N=277,344$) that takes the value of 1 if individual $i$ purchased at least one package of antidepressants prescribed by a practitioner in 2011. On the town level, high values of the mean of $A_{i}$ are clustered in space (Figure 1) and are positively correlated with log-transformed town population in a univariate linear regression ($\beta=0.038, p=0.041$), and negatively with average income (log-transformed, $\beta=-0.198, p=8.84e^{-08}$) (see Supporting Information 3).  A total of 10,291 individuals (3.7\%) have taken antidepressants in our small-town sample in 2011. The probability of using antidepressants is slightly lower for individuals who have above-medium $LC_{i}$ and is significantly lower for those individuals who have above-medium $SD_{i}$ (Figure 2B).

Next, we quantify the days of therapy (DOT) of those patients who take antidepressants in year 2011. To obtain the days of therapy $Z_{i,t}$, for each antidepressant type, we multiply the volume of  antidepressant packages purchased in year $t$ by individual $i$ with the per-package DOT value (as included in our data), and add up these products. Following these patients over the 2011-2015 period, we observe a decreasing trend (Figure 2C), and a decreasing correlation of $Z_{i,t}$ (Figure 2D).

\subsection*{Analysis}
We first estimate the likelihood of antidepressant use with a linear probability model, in which $A_{i}$ is predicted with an ordinary least squares (OLS) regression including the main explanatory variables and further controls:

\begin{equation}
P(A_{i}=1) = \alpha 
+ \beta_{1} LC_{i}
+ \beta_{2} {SD_{i}}
+ \beta_{3} ln \left (d_{i} \right )
+ \beta_{4} F_{i}^{r}
+ \beta \mathbf{X}_{i}
+ \beta \mathbf{S}_{h}
+ \mathbf{D}_{c},
\end{equation}
where $ln (d_{i})$ is the natural logarithm of ego network degree which has been found to impact mental health \cite{rosenquist2011social}, and $F_{i}^{r}$ is the concentration of connections within a radius (here we apply $r=50km$) that has been used to measure isolation from distant opportunities \cite{bailey2018social}. Age and Gender are important individual characteristics that correlate with mental health \cite{rosenquist2011social} and are denoted here by $\mathbf{X}_{i}$. A collection of settlement-level variables, including income or unemployment that we cannot control for on the individual level due to lack of data, are denoted by $\mathbf{S}_{h}$. $\mathbf{D}_{c}$ denotes county dummies that aims to capture the spatial correlation of $A_i$. The linear probability model allows us to analyse the interplay between $LC_i$ and $SD_i$ in explaining $A_i$. We use standardized values of all variables.

We confirm that female gender ($\beta=0.020, p<2e^{-16}$), and age ($\beta=0.020, p<2e^{-16}$) 
are significant predictors of mental health in our sample, in line with previous findings \cite{rosenquist2011social}. The economic status of the environment also plays a role as antidepressant use is less likely in towns where average income is relatively high ($\beta=-0.001, p=0.031$) and tends to be significantly higher in towns where unemployment rate is high ($\beta=0.001, p=0.026$). We find that individuals with more reported friends are less likely to take antidepressants ($p<2e^{-16}$) (Figure 3A). Having a geographically bounded social network, unlike in the case of economic development\cite{bailey2018social}, does not correlate significantly with mental health problems ($p=0.147$). The regression table of the estimation is reported in Supporting Information 5 where we also introduce and discuss a logistic regression specification; the logistic regression model explains the variation of $A_{i}$ with $AUC=0.72$ (see Supporting Information 6).


\begin{figure*}[!t]
\centering\includegraphics[width=\textwidth]{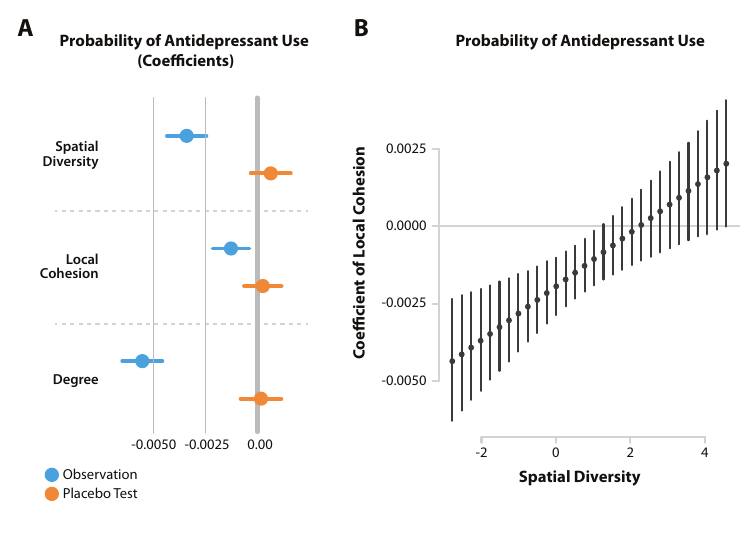}
\caption{Local Cohesion and Spatial Diversity of social networks can predict antidepressant use. Coefficients of variables standardized to $\zeta$-scores and 95\% confidence intervals are plotted. \textbf{A} Spatial Social Capital predicts the probability of antidepressant use compared to randomized outcome (placebo). \textbf{B} Local Cohesion looses significance as Spatial Diversity increases. 
}
\label{fig}
\end{figure*}

We find that $SD_{i}$ has a stronger negative association ($p=2.4e^{-12}$) with $A_i$ than $LC_{i}$ does ($p=0.005$) but weaker than the relationship with the number of social connections ($p<2e^{-16}$) (Figure 3A). This finding that the relationship between mental health and bridging social capital is stronger than with bonding social capital, which has been previously thought to dominate the maintenance of mental balance, is robust across various forms of $C_{i}$ normalization, using a narrower set of antidepressants, narrowing down the sample to individuals who have at least 10 friends, and using a logistic regression framework. Results do not change if we control for the antidepressant use of social contacts \cite{rosenquist2011social} ($p=2.82e^{-12}$) or for the distance to psychiatric centres \cite{tadmon2023differential} ($p=0.422$). (Robustness is reported in Supporting Information 7.) The Placebo test is run on the identical sample in which the variable $A_i$ is randomly reshuffled. None of the coefficients in the Placebo test are significant at the 5\% level.


In the next step, we investigate the interplay between Local Cohesion and Spatial Diversity to understand whether bonding and bridging complement each other for mental health as they do in the case of economic opportunities \cite{aral2016future}. We therefore estimate the previous equation with an additional interaction term between $LC_i$ and $SD_i$. The positive coefficient of the interaction term is reported in Supporting Information 9 and suggests that bridging and bonding social capital are not complementary in the case of mental health. Instead, Spatial Diversity mitigates the relationship between Local Cohesion and Antidepressant Use.

We therefore estimate the marginal effect of $LC_{i}$ on $A_i$ at levels of $SD_i$, a technique that provides detailed information about their interplay. We find that the negative relationship between $A_i$ and $LC_{i}$ is stronger at low levels of $SD_{i}$ than at medium levels and becomes insignificant at high levels of $SD_{i}$ (Figure 3B). The finding is robust against alternative normalizations of $C_i$ (see Supporting Information 8). This evidence suggests that the access to diverse groups (and bridging social capital) reduces the importance of cohesive local groups (and bonding social capital) for maintaining mental health. 

Finally, we analyze the dynamics of Days of Therapy ($Z_{i,t}$) of those patients who took antidepressants in 2011 (N=10,291). Since $Z_{i,t}$ is strongly correlated across subsequent years (reported in Figure 2D in the main text), controlling for the value of $Z_{i,2011}$ enables us to evaluate the forms of Spatial Social Capital in mitigating the dose of antidepressants in subsequent years, conditional on current use of antidepressants. We estimate the following equation with Ordinary Least Squares regression:

\begin{equation}
Z_{i,t} = \alpha 
+ \beta_{1} Z_{i,2011}
+ \beta_{2} \hat{A_h}
+ \beta_{3} LC_{i}
+ \beta_{4} {SD_{i}}
+ \beta_{5} ln \left (d_{i} \right )
+ \beta_{6} F_{i}^{r}
+ \beta \mathbf{X}_{i}
+ \beta \mathbf{S}_{h}
+ \mathbf{D}_{c}
+ \epsilon_{i,t},
\end{equation}
where $\hat{A_h}$ is the predicted probability that the individual living in town $h$ takes antidepressants as a function of her distance to the nearest psychiatric center, which has been argued to increase the access to treatment \cite{mcclellan1994does, tadmon2023differential}, and the ratio of antidepressant users in the region of the town that captures spatial inequalities of antidepressant use (as reported in Figure 1B). The inclusion of $\hat{A_h}$ helps us mitigate the selection bias of unequal access to antidepressants. More details of $\hat{A_h}$ prediction can be found in Supporting Information 9. The other co-variates in Equation 4 are identical to the ones in Equation 3.

\begin{table}[!b] \centering 
  \caption{Days of Therapy for patients who took antidepressants in 2011, Ordinary Last Squares regression. Not reported variables are town-level income and unemployment, and friends within 50km. Standard errors are in parantheses. $^{***,**,*}$ denote significance at the 1, 5, 10 percent level.} 
  \label{} 
\begin{tabular}{@{\extracolsep{5pt}}lccccc} 
\\[-1.8ex]\hline 
\hline \\[-1.8ex] 
\\[-1.8ex] & (1) & (2) & (3) & (4) & (5)\\ 
\\[-1.8ex] & $Z_{i,2013}$ & $Z_{i,2013}$ & $Z_{i,2013}$ & $Z_{i,2013}$ & $Z_{i,2015}$\\ 
\hline \\[-1.8ex] 
 $Z_{i,2011}$ & 1.394$^{***}$ & 1.393$^{***}$ & 1.393$^{***}$ & 1.393$^{***}$ & 1.221$^{***}$ \\ 
  & (0.020) & (0.020) & (0.020) & (0.020) & (0.022) \\ 
 Male$_i$ & $-$0.251$^{***}$ & $-$0.247$^{***}$ & $-$0.248$^{***}$ & $-$0.248$^{***}$ & $-$0.305$^{***}$ \\ 
  & (0.048) & (0.050) & (0.050) & (0.050) & (0.053) \\ 
 Age$_i$ & 0.335$^{***}$ & 0.334$^{***}$ & 0.333$^{***}$ & 0.334$^{***}$ & 0.312$^{***}$ \\ 
  & (0.026) & (0.027) & (0.027) & (0.027) & (0.029) \\ 
 $SD_i$ &  & $-$0.071$^{**}$ & $-$0.068$^{**}$ & $-$0.122$^{***}$ & $-$0.138$^{***}$ \\ 
  &  & (0.031) & (0.031) & (0.041) & (0.044) \\ 
 $LC_i$ &  & $-$0.023 & $-$0.022 & $-$0.026 & $-$0.021 \\ 
  &  & (0.027) & (0.027) & (0.027) & (0.029) \\ 
 $ln(d_i)$ &  & $-$0.057$^{*}$ & $-$0.055$^{*}$ & $-$0.054$^{*}$ & $-$0.039 \\ 
  &  & (0.030) & (0.030) & (0.030) & (0.032) \\ 
 $\hat{A_{h}}$ &  &  & 9.021$^{*}$ & 9.315$^{*}$ & 15.087$^{***}$ \\ 
  &  &  & (5.165) & (5.165) & (5.505) \\ 
 $SD_i$ × Male$_i$ &  &  &  & 0.010 & 0.085 \\ 
  &  &  &  & (0.053) & (0.056) \\ 
 $SD_i$ × Age$_i$ &  &  &  & 0.072$^{***}$ & 0.066$^{**}$ \\ 
  &  &  &  & (0.028) & (0.030) \\ 
 Constant & $-$4.016$^{***}$ & $-$4.064$^{***}$ & $-$4.409$^{***}$ & $-$4.409$^{***}$ & $-$3.816$^{***}$ \\ 
  & (0.172) & (0.176) & (0.265) & (0.265) & (0.282) \\ 
\hline \\[-1.8ex] 
Observations & 10,291 & 9,769 & 9,769 & 9,769 & 9,769 \\ 
R$^{2}$ & 0.355 & 0.355 & 0.355 & 0.355 & 0.277 \\ 
Adjusted R$^{2}$ & 0.354 & 0.353 & 0.353 & 0.353 & 0.274 \\ 
\hline\hline \\[-1.8ex] 
\end{tabular} 
\end{table} 

Table 1 presents regression results in a step-wise manner for $t \in \big\{2013, 2015\}$. Known predictors of mental health are introduced in Model (1) that confirms relatively faster recovery for male and relatively young patients. Adding social capital measures in Model (2), we find that $SD_i$ has a significant negative correlation with $Z_{i,2013}$, that $LC_{i}$ is not significant, and that the significance of Degree above the 5\% level. This finding suggests that bridging social capital has a mitigation effect on antidepressant use. The findings do not change when we introduce $\hat{A_h}$ in Model (3). We then test the heterogeneous impact of $SD_i$ on vulnerable population in Model (4) by including its interaction with the Age and Male variables and find that younger individuals benefit more from bridging social capital but there is no significant difference across gender in this respect. All relationships hold when $Z_{i,2015}$ is used as dependent variable in Model (5). Supporting Information 10 contains results for $t \in \big\{2012, 2013, 2014, 2015\}$ and for an alternative dependent variable $\Delta Z_{i,t}$ that captures the change of Days of Therapy between year 2011 and $t$.

\section*{Discussion}
In sum, we use nationwide data sets regarding antidepressant prescriptions and online social networks to analyze the relationship between social network structure and mental healthcare. Our data do not allow us to infer a causal relation. Distress, depressive symptoms, and anxiety are known to hinder social relations. Yet, the large scale of the data signals a previously undocumented correlation between the spatial diversity of social connections and mental health. This evidence suggests that bridging social capital and weak ties are important for mental health.

The theoretical importance of such bridging social capital has been previously highlighted by scholars who argue that engagement in more communities creates opportunity to develop connections of various functions, reduces isolation, and consequently, helps maintain mental balance \cite{fiori2006,salehi2019bonding}. 
Bridging social capital is extremely important for members of geographically isolated communities - where the lack of outside connections can make bonding social capital and cohesive social networks possibly even harmful for mental health by placing too much control on the individual \cite{almedom2005social, mitchell2002social} or by isolating the individual in an unhealthy social environment \cite{harpham2002measuring}. Unlike the case of economic advantage, where diversity in social networks is thought to complement network cohesion \cite{aral2016future}, we demonstrate that bridging social capital across distant groups is a substitute to local social bonding and can help patients with mental disorders to reduce the intensity of their treatment.

We also find that diverse connections to distant communities is a greater help for younger than for older individuals but does not significantly differ across residents of relatively rich versus relatively poor towns. This result might signal different usage of online social networks across generations such that younger cohorts can absorb and thus benefit more diverse information than older cohorts \cite{pfeil2009age, singleton2016online}. Yet, our data do not allow us to rule out alternative explanations. For example, younger individuals might be more active in establishing bridges across distant communities than older individuals, a mechanism that would require the analysis of dynamic social networks that we do not have. Individual-level information on income would also help us better understand how bridging ties can help disadvantaged and prospective groups. Such future research is important to inform health policy about potential social networking tools to help patients with mental disorders.

\section*{Materials and Methods}

\subsection*{Data Sources}

The anonymized administrative data on antidepressant purchases has been provided by the Hungarian National Healthcare Service Centre. The data cover individuals who had at least one purchase of antidepressants or antibiotics over the 2011-2015 period. We use data on the purchase of medications in the ATC (Anatomical Therapeutic Chemical) group N06A (antidepressants) that were purchased through pharmacies; thus, hospital care is excluded. The exact type and amount of the medication purchased is known. We generate annual indicators of antidepressant use and fill in zero antidepressant prescription values for each individual for whom we do not observe antidepressant purchase in a given year. We obtain the gender, date of birth, and home town from inpatient (hospital) and outpatient specialist care records of the patients included in the medication use data, covering years 2011-2015. The data consist of 9,017,323 individuals (out of the total almost 10 Million Hungarian citizens).

The OSN data covers information from individual profiles that were publicly available on the International Who Is Who (iWiW) website. The platform was functioning from 2002 to 2014 and the entire data set with basic user information (gender, date of birth, home town) and connection data (establishment of friendship ties). The data include 2.7 Million users who reported more than 300 Million friendship ties by 2011. Further information about the data sources can be found in Supporting Information 1.

\subsection*{Methods}
The data of antidepressant purchases were restricted to settlements with at most 20,000 inhabitants (4,728,045 individuals) but not less than 5,000 inhabitants. Both the antidepressant use data set and the social media data set were restricted to those date of birth -- town -- gender cells which have a single observation (2,163,158 individuals). Finally, the restricted files have been matched by the date of birth -- town -- gender characteristics, resulting in 298,441 matched observations. Supporting Information 2 contains a discussion on the estimated probability of failure of the matching process.

\section*{Ethics Statement}
The Scientific and Research Ethics Committee of the Hungarian Medical Science Council (ETT TUKEB, represented by Dr. Tamás Kardon, tukeb@bm.gov.hu) has provided ethical approval to match the antidepressant and social media datasets (Document ID: IV/5449-2/2021/EKU). All data are stored in an anonymized manner and individual patients have not been identified. 

\section*{Data and Code Availability}
The code that produces all results will be published in a public repository along with the final version of the paper. A data table that contains the necessary variables to reproduce the results will be also published. The published data will, however, not include personal information that would allow identification of individuals. The original iWiW dataset is protected with an NDA but can be accessed after authorization at the Data Bank of the Centre for Economic and Regional Studies. Further information about this can be requested from the corresponding author at lengyel.balazs@krtk.hun-ren.hu. The original antidepressant data cannot be accessed, since its' use has been restricted to the members of the research team.

\section*{Conflict of Interest}
The authors declare no conflict of interests.

\section*{Acknowledgement}
The work of Balázs Lengyel was financially supported by Hungarian National Scientific
Fund (OTKA K 138970). The authors acknowledge the help of Szabolcs Tóth-Zs. who contributed to the creation of the figures. The authors thank the comments from Péter Elek and from Srebrenka Letina.

\bibliography{scibib}

\begin{thebibliography}{10}

\bibitem{putnam2000bowling}
R.~D. Putnam, {\it Bowling alone: The collapse and revival of American community\/} (Simon and Schuster, 2000).

\bibitem{coleman1988social}
J.~S. Coleman, {\it American Journal of Sociology\/} {\bf 94}, S95 (1988).

\bibitem{mckenzie2002social}
K.~McKenzie, R.~Whitley, S.~Weich, {\it The British Journal of Psychiatry\/} {\bf 181}, 280 (2002).

\bibitem{henderson2003social}
S.~Henderson, H.~Whiteford, {\it The Lancet\/} {\bf 362}, 505 (2003).

\bibitem{almedom2005social}
A.~M. Almedom, {\it Social Science \& Medicine\/} {\bf 61}, 943 (2005).

\bibitem{rosenquist2011social}
J.~N. Rosenquist, J.~H. Fowler, N.~A. Christakis, {\it Molecular Psychiatry\/} {\bf 16}, 273 (2011).

\bibitem{elmer2020students}
T.~Elmer, K.~Mepham, C.~Stadtfeld, {\it PloS one\/} {\bf 15}, e0236337 (2020).

\bibitem{santini2020social}
Z.~I. Santini, {\it et~al.\/}, {\it The Lancet Public Health\/} {\bf 5}, e62 (2020).

\bibitem{smith2008}
K.~P. Smith, N.~A. Christakis, {\it Annual Review of Sociology\/} {\bf 34}, 405 (2008).

\bibitem{greenblatt1982social}
M.~Greenblatt, R.~M. Becerra, E.~Serafetinides, {\it The American Journal of Psychiatry\/}  (1982).

\bibitem{kawachi2001social}
I.~Kawachi, L.~F. Berkman, {\it Journal of Urban Health\/} {\bf 78}, 458 (2001).

\bibitem{lin2017building}
N.~Lin, {\it Social Capital\/} pp. 3--28 (2017).

\bibitem{fiori2006}
K.~L. Fiori, T.~C. Antonucci, K.~S. Cortina, {\it The Journals of Gerontology Series B: Psychological Sciences and Social Sciences\/} {\bf 61}, P25 (2006).

\bibitem{salehi2019bonding}
A.~Salehi, C.~Ehrlich, E.~Kendall, A.~Sav, {\it Journal of Mental Health\/} {\bf 28}, 331 (2019).

\bibitem{granovetter1973strength}
M.~S. Granovetter, {\it American Journal of Sociology\/} {\bf 78}, 1360 (1973).

\bibitem{eagle2010network}
N.~Eagle, M.~Macy, R.~Claxton, {\it Science\/} {\bf 328}, 1029 (2010).

\bibitem{chetty2022social}
R.~Chetty, {\it et~al.\/}, {\it Nature\/} {\bf 608}, 108 (2022).

\bibitem{thomson2022income}
R.~M. Thomson, {\it et~al.\/}, {\it The Lancet Public Health\/} {\bf 7}, e515 (2022).

\bibitem{dunbar1998social}
R.~I. Dunbar, {\it Evolutionary Anthropology: Issues, News, and Reviews: Issues, News, and Reviews\/} {\bf 6}, 178 (1998).

\bibitem{glaeser2002economic}
E.~L. Glaeser, D.~Laibson, B.~Sacerdote, {\it The Economic Journal\/} {\bf 112}, F437 (2002).

\bibitem{borgatti2009network}
S.~P. Borgatti, A.~Mehra, D.~J. Brass, G.~Labianca, {\it Science\/} {\bf 323}, 892 (2009).

\bibitem{liben2005geographic}
D.~Liben-Nowell, J.~Novak, R.~Kumar, P.~Raghavan, A.~Tomkins, {\it Proceedings of the National Academy of Sciences\/} {\bf 102}, 11623 (2005).

\bibitem{lambiotte2008geographical}
R.~Lambiotte, {\it et~al.\/}, {\it Physica A: Statistical Mechanics and its Applications\/} {\bf 387}, 5317 (2008).

\bibitem{lengyel2015geographies}
B.~Lengyel, A.~Varga, B.~S{\'a}gv{\'a}ri, {\'A}.~Jakobi, J.~Kert{\'e}sz, {\it PloS one\/} {\bf 10}, e0137248 (2015).

\bibitem{pulkki2012}
L.~Pulkki-R{\aa}back, {\it et~al.\/}, {\it BMC Public Health\/} {\bf 12}, 1 (2012).

\bibitem{sinokki2009}
M.~Sinokki, {\it et~al.\/}, {\it Journal of Affective Disorders\/} {\bf 115}, 36 (2009).

\bibitem{gardarsdottir2007}
H.~Gardarsdottir, E.~R. Heerdink, L.~van Dijk, A.~C. Egberts, {\it Journal of Affective Disorders\/} {\bf 98}, 109 (2007).

\bibitem{oksanen2011}
T.~Oksanen, {\it et~al.\/}, {\it Epidemiology\/} {\bf 22}, 553 (2011).

\bibitem{wachs2019social}
J.~Wachs, T.~Yasseri, B.~Lengyel, J.~Kert{\'e}sz, {\it Royal Society Open Science\/} {\bf 6}, 182103 (2019).

\bibitem{toth2021inequality}
G.~T{\'o}th, {\it et~al.\/}, {\it Nature Communications\/} {\bf 12}, 1143 (2021).

\bibitem{bailey2018social}
M.~Bailey, R.~Cao, T.~Kuchler, J.~Stroebel, A.~Wong, {\it Journal of Economic Perspectives\/} {\bf 32}, 259 (2018).

\bibitem{steinfield2008social}
C.~Steinfield, N.~B. Ellison, C.~Lampe, {\it Journal of Applied Developmental Psychology\/} {\bf 29}, 434 (2008).

\bibitem{shakya2017association}
H.~B. Shakya, N.~A. Christakis, {\it American Journal of Epidemiology\/} {\bf 185}, 203 (2017).

\bibitem{braghieri2022social}
L.~Braghieri, R.~Levy, A.~Makarin, {\it American Economic Review\/} {\bf 112}, 3660 (2022).

\bibitem{neal2017small}
Z.~P. Neal, {\it Network Science\/} {\bf 5}, 30 (2017).

\bibitem{tadmon2023differential}
D.~Tadmon, P.~S. Bearman, {\it Proceedings of the National Academy of Sciences\/} {\bf 120}, e2301304120 (2023).

\bibitem{aral2016future}
S.~Aral, {\it American Journal of Sociology\/} {\bf 121}, 1931 (2016).

\bibitem{mcclellan1994does}
M.~McClellan, B.~J. McNeil, J.~P. Newhouse, {\it Jama\/} {\bf 272}, 859 (1994).

\bibitem{mitchell2002social}
C.~U. Mitchell, M.~LaGory, {\it City \& Community\/} {\bf 1}, 199 (2002).

\bibitem{harpham2002measuring}
T.~Harpham, E.~Grant, E.~Thomas, {\it Health Policy and Planning\/} {\bf 17}, 106 (2002).

\bibitem{pfeil2009age}
U.~Pfeil, R.~Arjan, P.~Zaphiris, {\it Computers in Human Behavior\/} {\bf 25}, 643 (2009).

\bibitem{singleton2016online}
A.~Singleton, P.~Abeles, I.~C. Smith, {\it Computers in Human Behavior\/} {\bf 61}, 394 (2016).

\end{thebibliography}

\bibliographystyle{Science}

\clearpage

\renewcommand{\thefigure}{S\arabic{figure}}
\renewcommand{\thetable}{S\arabic{table}} 
\renewcommand{\theequation}{S\arabic{equation}} 
\setcounter{figure}{0}
\setcounter{table}{0}
\setcounter{equation}{0}

\section*{Supporting Information}
\section*{Supporting Information 1: Data Sources}

We use anonymized administrative data on antidepressant purchases for the entire population of Hungary, provided by the Hungarian National Healthcare Service Centre. The sample consists of individuals who had at least one purchase of antidepressants or antibiotics for systemic use over the 2011-2015 period.\footnote{Based on an administrative data set on a random 50\% of the Hungarian population (data available at the Data Bank of the  Centre for Economic and Regional Studies), 89.7\% of the population had at least one purchase of antidepressants or antibiotics for systemic use over 2011-2015.} The pharmaceutical records we use in this paper show data on the purchase of medications in the ATC (Anatomical Therapeutic Chemical) group N06A (antidepressants) that were purchased through pharmacies. The medication records relate only to the ambulatory setting and thus exclude hospital care. We know the exact type and amount of the medication purchased, hence days of therapy (DOT) can be calculated. We obtain the gender, date of birth, and settlement indicator from inpatient and outpatient care records covering years 2011-2015. These indicators are needed for matching the drug usage and our network (iWiW) data. Importantly, we then have the gender, date of birth, and settlement indicator for all individuals who purchased antidepressants or antibiotics for systemic use at least once, and had an outpatient or inpatient care event in 2011-2015, resulting in a data set of 9,017,323 (Hungary had a population of 9,9 million in 2011). We generate annual indicators of antidepressant use and fill in zero antidepressant prescription values for each individual for whom we do not observe antidepressant purchase in a given year. 

To capture Spatial Social Capital, we use data from a Hungarian online social network called iWiW. This website was the market leader social media platform used by nearly 40\% of the country’s population. iWiW was launched in 2002 and shortly became the most widely used online social network in Hungary. At its peak around 2008-2010, it was one of the most visited national websites reaching the majority of internet users of the country. During the first few years of operation, iWiW provided only basic functionalities, mostly built around finding present and former friends, classmates, colleagues, and looking through one’s acquaintance’s’ acquaintances. Later, photo upload, news-feed, messaging, applet to visualize connections and the ability to develop external applications was introduced to the service. However, iWiW failed to follow leading social media portals in terms of hosting the increasingly intense online communication. Instead, iWiW remained a phone-book type online collection of friends. 

Due to the increasing maintenance costs, low profitability and tough competition from Facebook, the site was closed down on June 30, 2014. Although the number of daily visitors begun to fall back significantly from 2011-2012, users rarely deleted their profiles: they just abandoned the service. In February 2013, the entire dataset of iWiW with basic user information (i.e., date of registration, gender, age, etc.) and  connection data (establishment of friendship ties) were made available for us for scientific research purposes. The self-reported location of users is available at the town level. We can analyze more than 300 million friendship ties the users have established by the end of 2011. Previous research demonstrated that geographical factors explain registration rates on the website \cite{lengyel2015geographies}, and found relation between structural measures of social capital and outcomes such as the prevalence of corruption in towns \cite{wachs2019social} and dynamics of income inequalities in towns \cite{toth2021inequality}.

\section*{Supporting Information 2: Matching of Social Media with Antidepressant Data}

We restrict the data of antidepressant purchases to settlements with at most 20,000 inhabitants (4,728,045 individuals), to reduce the probability of false matches between the drug use and iWiW data. Next, we exclude individuals who live in towns where population size is less than 5,000 inhabitants because the fraction of iWiW users fell sharply in these smallest towns \cite{toth2021inequality}. We restrict both the drug user file and the iWiW data file to those date of birth -- settlement -- gender cells which have a single observation (2,163,158 individuals). We then match the so restricted drug user file to the iWiW file, which is now a 1:1 matching, resulting in 298,441 matched observations.

The estimated probability of failure in the above matching procedure is low. If the drug user file covered the entire population then the restriction to those date of birth -- settlement -- gender cells which have a single observation would ensure that there are no false matches when matching the data with the iWiW file. However, we do not observe approximately 9\% of the population (for whom we do not have any medical records in 2011-2015). In a settlement of 20,000 inhabitants (the maximum in our restricted sample), we therefore do not observe on average 1,800 individuals. Assuming, for simplicity, that there are $65 \times 365 \times 2 = 47,450$ unique date of birth -- gender combinations which are evenly distributed in the population, then there is a 3.7\% probability that there is an individual among the 1,800 unobserved individuals who has the same date of birth -- gender combination as a specific individual in the observed pool ($0.037 = 1 - (47,449/47,450)^{1,800}$). This is the probability of a false match in the largest settlements in our sample, which is lower in smaller settlements.

Importantly, the mean annual days of therapy of antidepressants in 2012 does not differ substantially across the samples -- in the entire drug use file it is 10.66, in the sample restricted to settlements with at most 20,000 inhabitants it is 10.58, and in the sample restricted to cells with single birth -- settlement -- gender observations it is 10.47.

\section*{Supporting Information 3: Town-level Analysis}

Individuals in our sample reside in 409 towns. To demonstrate that this sample is not different from other towns in the country, we compare the distribution of the fraction of antidepressant users in sample towns and all other towns where we do not match the two data sources ($N=2219$). The two-sided Mann-Whitney U test is not significant ($p= 0.971$) indicating that the sample towns where we performed the matching between the iWiW and antidepressant data ($N=409$) are not significantly different from the rest of the country.

\begin{figure*}[!t]
\centering\includegraphics[width=0.7\textwidth]{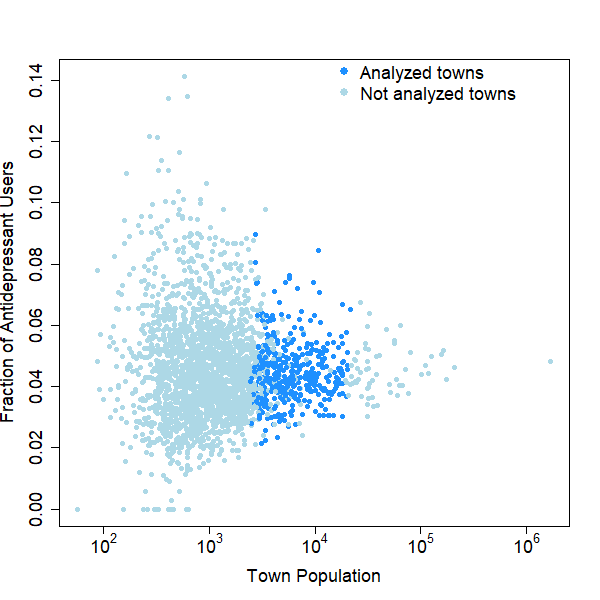}
\caption{Antidepressant usage is not correlated with town population.}
\label{figa1}
\end{figure*}

Next, we investigate whether town characteristics are correlated with the fraction of antidepressant users. We find no correlation between population and fraction of antidepressant users (both variables are log-transformed) in the total set of towns ($r=-0.031$) (Figure \ref{figa1}). Ordinary least squares regressions confirm that population becomes significantly associated with the fraction of antidepressant users only when further socio-economic characteristics of the sample towns, such as average income, unemployment rate, and distance to the closest country border are controlled for (Table \ref{tab_app_town}). We observe that antidepressants are taken by slightly more individuals in relatively larger towns. More importantly, antidepressant use is significantly lower in relatively wealthier towns. Therefore, we control for the socio-economic characteristics of towns in the individual-level regressions that we present in Figure 3 in the main text.

\begin{table}[!ht] \centering 
  \caption{Fraction of antidepressant users and town characteristics. Ordinary Least Squares regression.} 
  \label{tab_app_town} 
\begin{tabular}{@{\extracolsep{5pt}}lcccc} 
\\[-1.8ex]\hline 
\hline \\[-1.8ex] 
\\[-1.8ex] & \multicolumn{3}{c}{Fraction of Antidepressant Users (\%, log)} \\ 
 & M1 & M2 & M3 & M4\\ 
\hline \\[-1.8ex] 
Population (\#, log) & -0.001 & 0.017 & 0.038$^{**}$ & 0.038$^{**}$ \\ 
  & (0.000) & (0.019) & (0.019) & (0.018) \\ 
  & & & & \\ 
Average Income (HUF, log) & &  & $-$0.198$^{***}$ & $-$0.192$^{***}$ \\ 
  & &  & (0.036) & (0.036) \\ 
  & & & & \\ 
Unemployment Rate (\%, log) & &  &  & $-$0.071 \\ 
  & &  &  & (0.350) \\ 
  & & & & \\ 
Distance to Border (km, log) & &  &  & $-$0.029$^{***}$ \\ 
  & &  &  & (0.009) \\ 
  & & & & \\ 
 Constant & 0.048$^{***}$ & $-$3.274$^{***}$ & $-$0.807$^{*}$ & $-$0.772 \\ 
  & (0.001) & (0.163) & (0.479) & (0.478) \\ 
  & & & & \\ 
Observations & 2624 & 409 & 409 & 409 \\ 
R$^{2}$ & 0.001 & 0.002 & 0.070 & 0.092 \\ 
Adjusted R$^{2}$ & 0.000 & $-$0.001 & 0.065 & 0.083 \\ 
\hline \\[-1.8ex] 
\multicolumn{4}{l}{\textit{Note:} Standard errors in parentheses. $^{*}$p$<$0.1; $^{**}$p$<$0.05; $^{***}$p$<$0.01} \\ 
\end{tabular} 
\end{table} 

\section*{Supporting Information 4: Variables and Descriptive Statistics}

Research has consistently shown that close social relationships are beneficial for maintaining good mental health. Being surrounded by a community that one can rely on provides a sense of security and confidence, which can assist in managing anxiety. In the social capital literature, this idea is often operationalized through clustering, also known as transitivity concerning social relationships. Clustering refers to the proportion of closed triples in a social network. Higher transitivity values foster trust and facilitate the creation of cohesive communities. In our research, we extend the conventional clustering calculation by incorporating spatial considerations and, more specifically, examining the degree of social cohesion within one's immediate living environment. This concept is referred to as Local Cohesion and is denoted by $LC_{i_{h}}$, where $h$ refers to the hometown of the individual.

Another crucial variable that is rather new in the literature of empirical research on mental health is social network diversity, particularly in terms of its spatial dimension. Thus far, researches have not been able to explore individuals' weak connections and their potential correlations with mental health outcomes. While some studies have shown that a more diverse communication telephone network is associated with higher financial well-being \cite{eagle2010network}, the effects of social network diversity on mental health have not been extensively examined. In our study, we incorporate Spatial Diversity as a variable denoted by $SD_{i}$ into our regression analysis. 

\begin{figure*}[!t]
\centering\includegraphics[width=\textwidth]{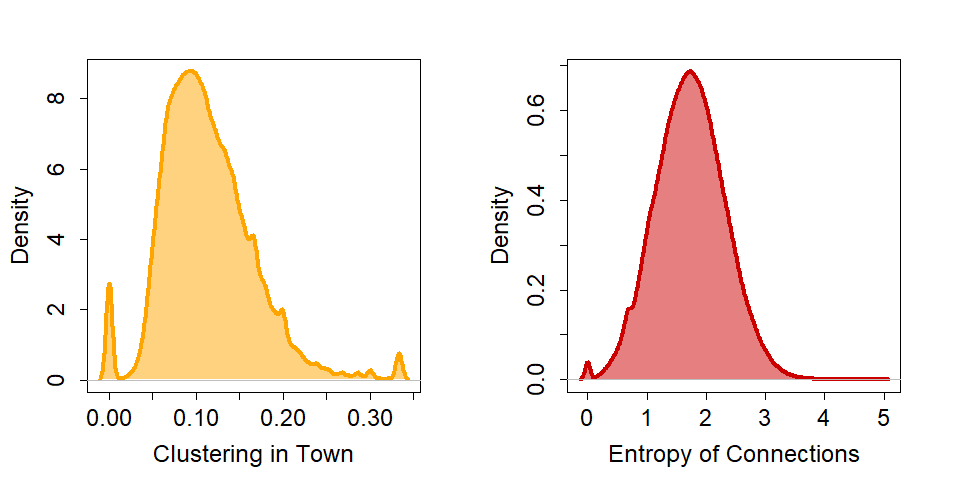}
\caption{Distribution of the non-normalized values of local clustering in the town of Ego and the entropy of Ego's connections across towns.}
\label{figa2}
\end{figure*}

To compute these indices, we first measure the local clustering coefficient of individuals in their friendship networks within their home town that is the basis of the Local Cohesion variable and the entropy of friendship ties across towns that is the basis of the Spatial Diversity variable. Both of these measures follow a distribution that is close to normal (Figure \ref{figa2}). Yet, these measures can be correlated with degree. Therefore, we apply two normalization methods that enables the comparison of the network structure among individuals with very different number of friends in their hometown and in other towns as well. 

\begin{figure*}[!ht]
\centering\includegraphics[width=\textwidth]{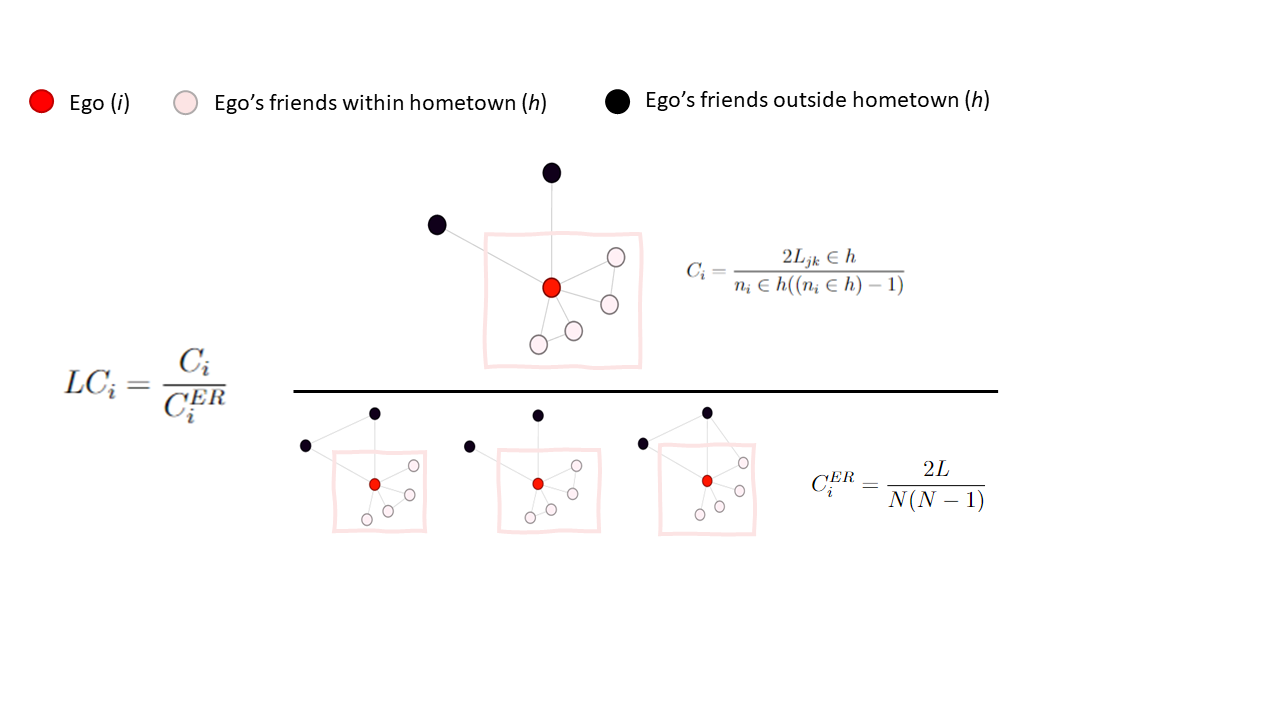}
\caption{Calculation of the Local Cohesion variable. The local clustering is compared to randomly rewired networks that keep degree and ego network density but do not consider town borders.}
\label{figa3}
\end{figure*}

Local Cohesion is quantified by dividing the local clustering of individuals within their home-town network by the local clustering of a series of randomly rewired networks that keep the the degree of ego and the number of links in the ego network (Figure \ref{figa3}). This procedure randomizes links in the network of individuals without considering town borders. Therefore, Local Cohesion captures the intensity of triadic closure within hometown such that it is not correlated with degree. 

\begin{figure*}[!ht]
\centering\includegraphics[width=\textwidth]{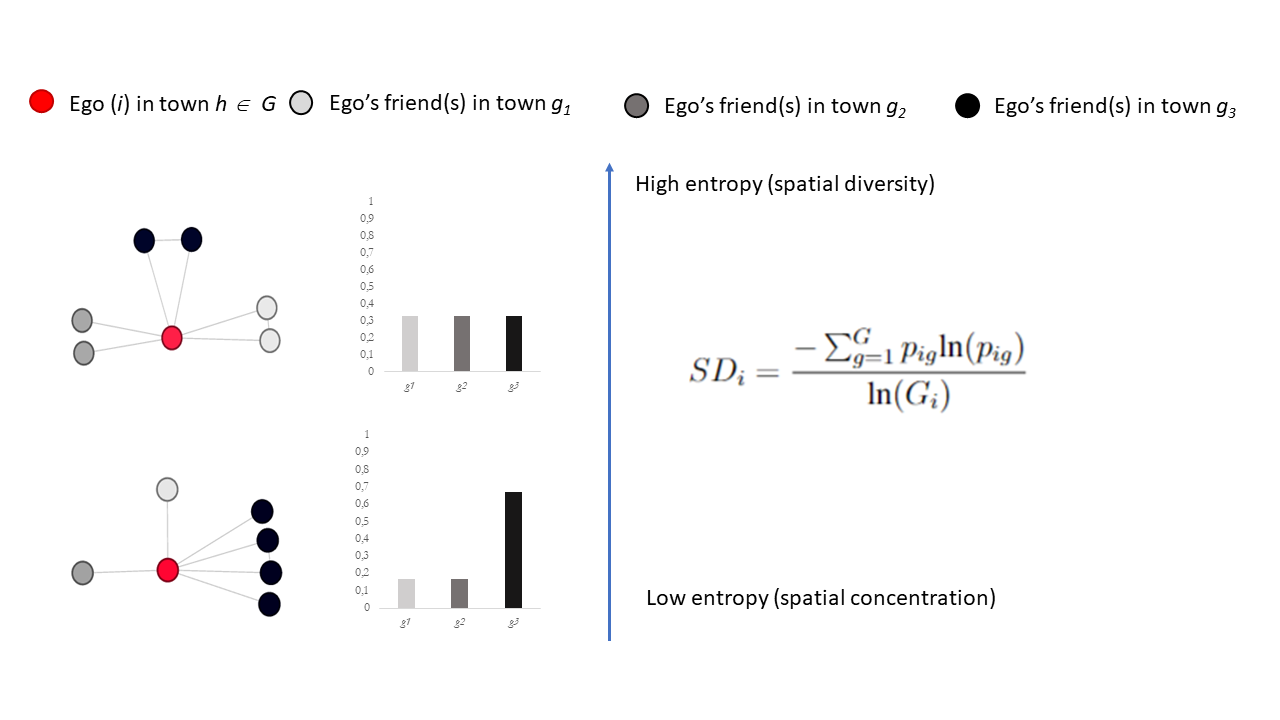}
\caption{Calculation of the Spatial Diversity variable. The entropy of connections across towns is divided by the number of towns the individual has access to.}
\label{figa4}
\end{figure*}

Next, Spatial Diversity is calculated by dividing the entropy of social connections across towns by the number of towns that the individual has access to (Figure \ref{figa4}). This procedure is the standard method to compar diversity of high degree and low degree individuals \cite{eagle2010network}.

\begin{figure*}[!t]
\centering\includegraphics[width=\textwidth]{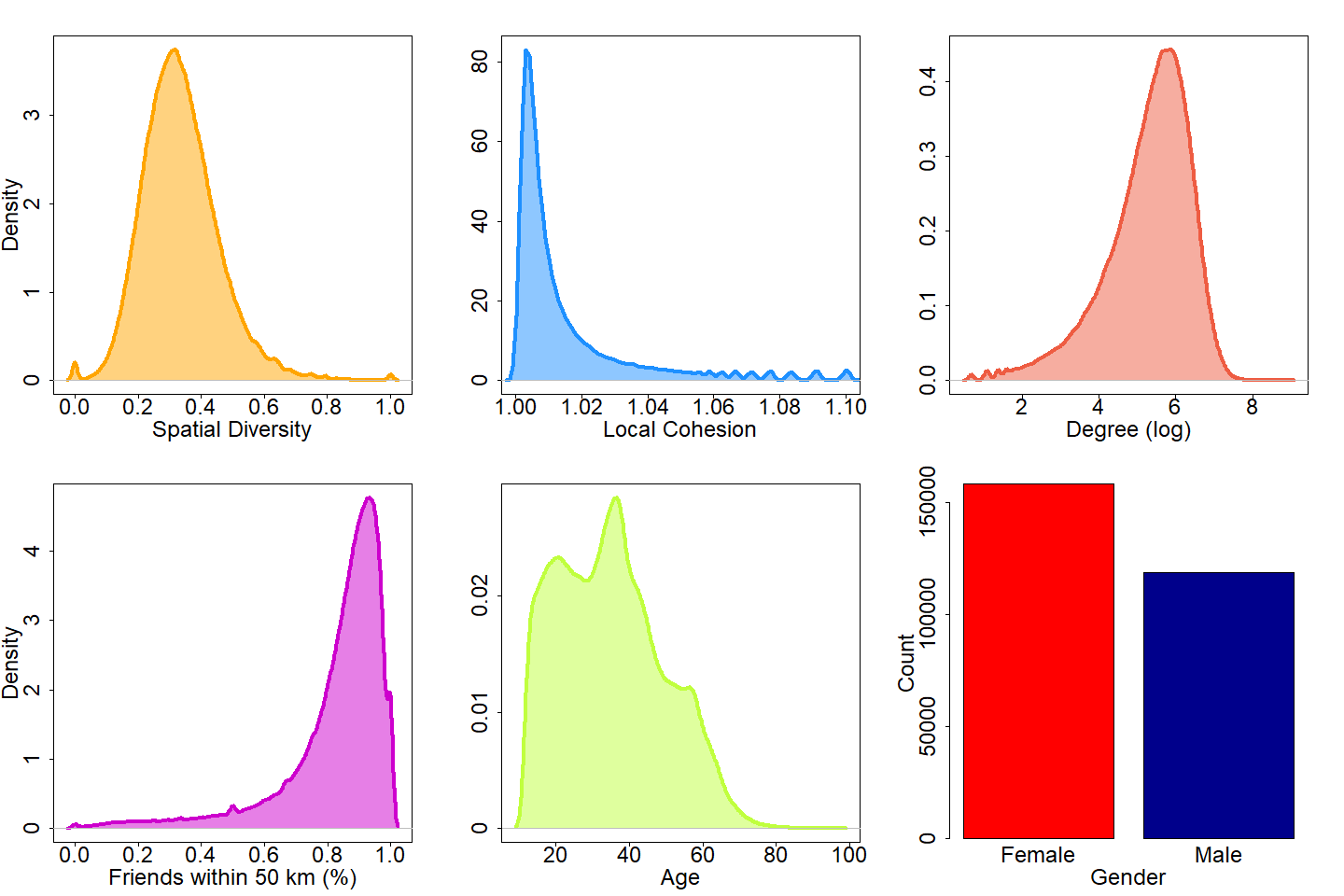}
\caption{Distributions of individual-level variables.}
\label{figa5}
\end{figure*}

The above normalization procedure does not induce a radical change to the entropy distribution; Spatial Diversity is close to normal. However, the distribution of Local Cohesion becomes right-skewed indicating that local clustering in home-town are similar to randomly rewired networks for most individuals in the sample (Figure \ref{figa5}).

The individual-level analysis contains further network variables and demographic controllers. Degree measures the log-transformed number of friends on iWiW that has been repeatedly shown to correlate with mental health \cite{shakya2017association}. Social networks in remote areas in Hungary are frequently concentrated within close geographical proximity \cite{lengyel2015geographies} that can hinder access to diverse information \cite{bailey2018social}. Thus, we include the fraction of friends within a specified radius, denoted by $r$. In the estimation model, $r$ has been set to 50km that reflects meaningful spatial scales in Hungary.  To capture the relationship between the number of friends and antidepressant use, we take the logarithm of the former variable. 

Age and gender are also included as important determinants of mental health. Figure \ref{figa5} depicts the distribution of these variables and Table \ref{tab_app_descr} presents their descriptive statistics.

Mental disorders have been shown to propagate in social networks through peer effects \cite{rosenquist2011social}. Therefore, we create a binary variable ``Friend taking Antidepressant'' that takes the value of 1 if the individual has at least one friend who takes antidepressants and is 0 otherwise. The median value of of this indicator is 1, which means that most of the individuals in our sample know at least one person who takes antidepressants (Table \ref{tab_app_descr}). Unlike in previous studies, we now investigate a network where weak ties are abundant, probably channeling different mechanisms of mental disorder peer effects than strong ties. Therefore, we use this variable only in robustness checks. 

\begin{table}[ht]
\centering
\caption{Descriptive statistics of individual-level  variables}\label{tab_app_descr}
\begin{tabular}{rrrrrrrrrrrrrr}
  \hline
Variables & N & Mean & SD & Median & Min & Max\\ 
  \hline
Antidepressant Use & 277,279 & 0.03 & 0.19 & 0.00 & 0.00 & 1.00\\ 
  Spatial Diversity & 277,279 & 0.34 & 0.12 & 0.32 & 0.00 & 1.00\\ 
  Local Cohesion & 265,719 & 1.03 & 0.06 & 1.01 & 1.00 & 1.50\\ 
  Degree (log) & 277,279 & 5.27 & 1.10 & 5.48 & 0.69 & 8.87\\ 
  Friends within 50 km& 277,344 & 0.83 & 0.17 & 0.87 & 0.00 & 1.00\\ 
  Male & 277,279 & 0.43 & 0.49 & 0.00 & 0.00 & 1.00 \\ 
  Age & 277,279 & 34.94 & 14.28 & 34.00 & 12.00 & 96.00\\ 
  Friend taking Antidepressant & 275,877 & 0.66 & 0.47 & 1.00 & 0.00 & 1.00\\ \hline
\end{tabular}
\end{table}

\begin{table}[ht]
\centering
\caption{Pearson Correlation between individual-level explanatory variables}\label{tab_app_corr}
\begin{tabular}{rrrrrrrrr}
  \hline
 & & 1. & 2. & 3. & 4. & 5. & 6. \\ 
  \hline
1. & Spatial Diversity & 1.00 & & & & & \\ 
2. & Local Cohesion & 0.32 & 1.00 & & & & \\ 
3. & Degree & -0.22 & -0.55 & 1.00 & & & \\ 
4. & Friends within 50 km& -0.43 & -0.12 & -0.04 & 1.00 & & \\ 
5. & Male & 0.08 & 0.06 & -0.12 & -0.03 & 1.00 & \\ 
6. & Age & 0.10 & 0.06 & 0.07 & -0.06 & -0.04 & 1.00 \\ 
7. & Friend taking Antidepressant & -0.15 & -0.33 & 0.52 & 0.01 & -0.10 & 0.19 \\ 
   \hline
\end{tabular}
\end{table}

Pearson correlation values of independent variables are low in most cases (Table \ref{tab_app_corr}). Degree is negatively correlated with Local Cohesion implying that individuals who have many friends tend to have local clustering in their home town that is relatively similar to random networks. There is a positive correlation between Degree and Friend taking Antidepressant because the more friend one has the higher likelihood that on of them will use antidepressants. None of these correlations are strong enough to impose problems (variance inflation) of severe multicollinearity.

\section*{Supporting Information 5: The Probability of Antidepressant Use: Estimation Results}

To examine the relationship between mental health and Spatial Social Capital, we utilize multivariate regression analysis. Our binary dependent variable, $A_i$ is  equal to 1 if the individual purchased a prescribed antidepressant in 2012, serving as a useful proxy for mental health problems. Although the exact form of the latent variable is unknown, regressions using drug use as a proxy provide an adequate approximation of the probability of depression. The linear probability model (LPM) has increasing popularity in social sciences because it allows for a straightforward interpretation of the interaction term between explanatory variables that is important for our line of argument. However, using the LPM requires to assume a linear relationship between social network structure and the probability of antidepressant use that we cannot test. Thus, we also employ a logistic regression estimation framework as a robustness check.


Besides the LPM regression that is specified in Equation 3 in the main text, we also estimate a logistic regression model as follows:

\begin{equation}
ln \left (  \frac{P(A_{i}=1)}{1-P(A_{i}=1)}  \right ) = \alpha 
+ \beta_{1} LC_{i}
+ \beta_{2} {SD_{i}}
+ \beta_{3} ln \left (d_{i} \right )
+ \beta_{4} F_{i}^{r}
+ \beta \mathbf{X}_{i}
+ \beta \mathbf{S}_{h}
+ \mathbf{D}_{c},
\end{equation}

\def\sym#1{\ifmmode^{#1}\else\(^{#1}\)\fi}
\begin{table}[!h] \centering 
  \caption{Probability of Antidepressant Use. Linear probability model (LPM) and logistic regression.} 
  \label{tab_app_logit}
\begin{tabular}{@{\extracolsep{5pt}}lccc} 
\\[-1.8ex]\hline 
\hline \\[-1.8ex] 
\\[-1.8ex] & \textit{LPM} & \textit{Placebo} & \textit{Logistic} \\ 
\\[-1.8ex] & (1) & (2) & (3)\\ 
\hline \\[-1.8ex] 
 $SD_{i}$ & $-$0.003$^{***}$ & 0.001 & $-$0.029$^{**}$ \\ 
  & (0.000) & (0.000) & (0.014) \\ 
  & & & \\ 
 $LC_{i}$ & $-$0.001$^{***}$ & 0.000 & $-$0.012 \\ 
  & (0.000) & (0.000) & (0.013) \\ 
  & & & \\ 
 ln $\left (d_{i} \right )$ & $-$0.006$^{***}$ & 0.000 & $-$0.050$^{***}$ \\ 
  & (0.000) & (0.001) & (0.014) \\ 
  & & & \\ 
 $F_{i}^{50}$ & $-$0.001 & $-$0.000 & 0.004 \\ 
  & (0.000) & (0.000) & (0.014) \\ 
  & & & \\ 
 Male$_i$ & $-$0.020$^{***}$ & $-$0.000 & $-$0.628$^{***}$ \\ 
  & (0.001) & (0.001) & (0.023) \\ 
  & & & \\ 
 Age$_i$ & 0.026$^{***}$ & $-$0.000 & 0.688$^{***}$ \\ 
  & (0.000) & (0.000) & (0.010) \\ 
  & & & \\ 
 Income$_h$ & $-$0.001$^{**}$ & 0.001 & $-$0.049$^{**}$ \\ 
  & (0.001) & (0.001) & (0.020) \\ 
  & & & \\ 
 Unemployment$_h$ & 0.002$^{**}$ & 0.001 & 0.047$^{**}$ \\ 
  & (0.001) & (0.001) & (0.023) \\ 
  & & & \\ 
 Constant & 0.053$^{***}$ & 0.039$^{***}$ & $-$3.067$^{***}$ \\ 
  & (0.003) & (0.003) & (0.068) \\ 
  & & & \\ 
\hline
County  FE    &         Yes         &         Yes &         Yes        \\
\hline
N & 265,719 & 265,719 & 265,719 \\ 
R$^{2}$ & 0.022 & 0.000 &  \\ 
\hline\hline
\multicolumn{4}{l}{\textit{Note:} Standard errors in parentheses. $^{*}$p$<$0.1; $^{**}$p$<$0.05; $^{***}$p$<$0.01} \\ 
\end{tabular} 
\end{table} 

where $P(A_{i}=1)$ captures the probability that individual is an anti-depressant user, $ln (d_{i})$ is the natural logarithm of degree, and $F_{i}^{r}$ is the fraction of friends within a radius $r=50km$. Individual-level demographic variables Age and Gender are denoted by $\mathbf{X}_{i}$, a collection of settlement-level variables are denoted by $\mathbf{S}_{h}$, and $\mathbf{D}_{c}$ denotes county dummies. We include age and gender as control variables as these are expected to be related to the risk of depression, and also to the network characteristics. Moreover, we recognize that many individual-level characteristics contribute to mental balance, including income or unemployment. However, we do not have individual-level data on income and unemployment, therefore we control for them on the level of towns.

Table \ref{tab_app_logit} presents the estimation results. Besides the LPM findings reported in the main text, one can observe that the concentration of friendship ties has a no significant relationship with antidepressant use. As expected, men are significantly less likely to take antidepressants than women and the likelihood increases as the age grows; the findings are in line with previous findings \cite{rosenquist2011social}. Further, we find that individual antidepressant use is less likely in towns where average income is relatively high and more likely where the unemployment rate is relatively high. 

The Placebo test is run on the identical sample in which the variable $A_i$ is randomly reshuffled. None of the coefficients are significant, verifying that the results are not an artefact of chance.

Finally, we find that the logistic regression confirms the robustness of almost all variables. $LC_i$ is an exception as it's negative relationship with $A_i$ is not significant in the logistic regression. Yet, the relationship of $SD_i$ and degree remain significant.

\section*{Supporting Information 6: Predictive Accuracy}

The quality of LPM fit are usually very low. Therefore, we measure estimation precision of the logistic regression and focus on determining the confidence level at which we can differentiate between antidepressant users and non-users, referred to as ``model accuracy'', using our model specified in Equation S1 in Supporting Information 5. 

To assess the model accuracy, we calculate two key metrics: Sensitivity and Specificity. Sensitivity, or the true positive rate can be calculated from true positives (TP) and false negative (FN) by the formula $TP/(TP+FN)$. This indicator measures the proportion of individuals correctly classified as antidepressant user by the logit regression model. It reflects the probability of obtaining a positive test result when the individual is genuinely a user. Specificity, the true negative ratio, is calculated from true negatives (TN) and false positive (FP) by $TN/(FP+TN)$ and measures the proportion of individuals accurately classified as non-user. It signifies the probability of obtaining a negative test result when the individual is genuinely not using antidepressants.

\begin{figure*}[!t]
\centering\includegraphics[width=0.8\textwidth]{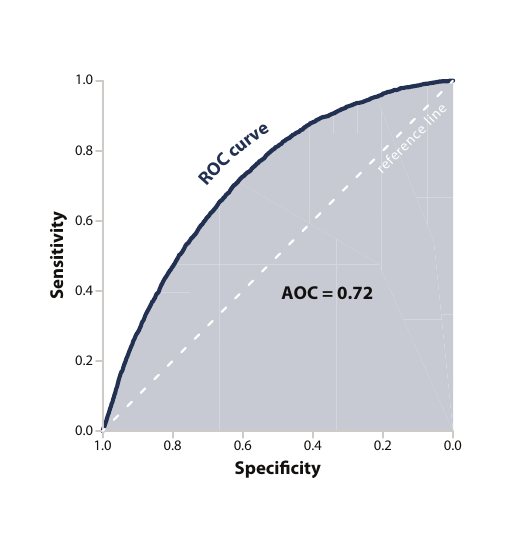}
\caption{Predictive accuracy of the logistic regression that estimates the Probability of Antidepressant Use.}
\label{figa7}
\end{figure*}

To visualize the model's ability to distinguish between antidepressant users and non-users, we construct a receiver operating characteristic (ROC) curve. This graphical representation plots the accuracy ratios of Sensitivity and Specificity. The ROC curve's 45-degree reference line denotes the accuracy achieved by random chance in classifying binary observations. A greater deviation of the ROC curve from the reference line indicates a higher discriminatory capability of our estimator. The area under the ROC curve provides a numerical measure of the predictive accuracy (AUC). Our logit regression model achieves a predictive accuracy of 72\% in distinguishing between antidepressant users and non-users.

Establishing a consensus on what constitutes a sufficiently high predictive accuracy value is challenging. Nonetheless, within the social sciences realm, our achievement of 72\% accuracy using social network variables and basic demographic indicators is considered notable. This finding underscores the potential of these variables to accurately discern between antidepressant users and non-users, contributing to advancements in mental health research.

\newpage
\section*{Supporting Information 7: Robustness}

{
\def\sym#1{\ifmmode^{#1}\else\(^{#1}\)\fi}
\begin{table}
    \centering
\caption{Robustness checks of Probability of Antidepressant Use estimation. Linear probability regressions.}\label{tab_app_robust1}
\resizebox{\textwidth}{!}{
\begin{tabular}{l*{7}{c}}
\hline\hline
            &\multicolumn{1}{c}{Model 1}&\multicolumn{1}{c}{Model 2}&\multicolumn{1}{c}{Model 3}&\multicolumn{1}{c}{Model 4}&\multicolumn{1}{c}{Model 5}&\multicolumn{1}{c}{Model 6} \\
\hline
            &\multicolumn{1}{c}{Baseline}&\multicolumn{1}{c}{Strict Definition (ICD-10)}&\multicolumn{1}{c}{d$\geq 10$}&\multicolumn{1}{c}{Friend}&\multicolumn{1}{c}{Psychiatry}&\multicolumn{1}{c}{Borders} \\
\hline
        &                     &                     &                     &                     &                     &                                        \\
$SD_{i}$      &     -0.0034\sym{***}&     -0.0034\sym{**} &     -0.0035\sym{***}&     -0.0033\sym{***}&     -0.0034\sym{***}&     -0.0034\sym{***}\\
            &     (0.000)         &     (0.000)         &     (0.000)         &     (0.000)         &     (0.000)         &     (0.000)     \\
[1em]
$LC_{i}$ &     -0.0012\sym{***}&     -0.0011\sym{**}        &     -0.0016\sym{***}&     -0.0010\sym{**}&     -0.0012\sym{***}&     -0.0013\sym{***} \\
            &     (0.000)         &     (0.000)         &     (0.000)         &     (0.000)         &     (0.000)         &     (0.000)         \\
[1em]
ln $\left (d_{i} \right )$         &     -0.0055\sym{***}&     -0.0049\sym{***}&     -0.0056\sym{***}&     -0.0070\sym{***}&     -0.0055\sym{***}&     -0.0056\sym{***} \\
                                     &     (0.000)         &     (0.000)         &     (0.000)         &     (0.000)         &     (0.000)         &     (0.000)     \\
[1em]
$F_{i}^{50}$          &      0.0006  &      0.0002         &      -0.0007         &      -0.0007         &      -0.0006  &      -0.0007     \\
            &     (0.000)         &     (0.000)         &     (0.000)         &     (0.000)         &     (0.000)         &     (0.000)          \\
[1em]
Male          &     -0.0201\sym{***}&     -0.0208\sym{***}&     -0.0200\sym{***}&     -0.0199\sym{***}&     -0.0201\sym{***}&     -0.0201\sym{***}    \\
            &     (0.000)         &     (0.000)         &     (0.000)         &     (0.000)         &     (0.000)         &     (0.000)                   \\
[1em]
Age         &      0.0261\sym{***}&      0.0245\sym{***}&      0.0260\sym{***}&      0.0255\sym{***}&      0.0261\sym{***}&      0.0261\sym{***} \\
            &     (0.000)         &     (0.000)         &     (0.000)         &     (0.000)         &     (0.000)         &     (0.000)                  \\
[1em]
Unemployment$_h$   &      0.0017\sym{***}  &      0.0019\sym{**} &      0.0017\sym{**}  &      0.0017\sym{**}  &      0.0017\sym{**}  &      0.0016\sym{**} \\
            &     (0.000)         &     (0.000)         &     (0.000)         &     (0.000)         &     (0.000)         &     (0.000)       \\
[1em]
Income$_h$ &     -0.0014\sym{**}&     -0.0009  &     -0.0013\sym{**}&     -0.0011\sym{*}&     -0.0013\sym{**}&     -0.0016\sym{**}\\
            &     (0.000)         &     (0.000)         &     (0.000)         &     (0.000)         &     (0.000)         &     (0.000)         \\
[1em]
Friend takes Antidep.       &                     &                     &                     &      0.0064\sym{***}&                     &                           \\
            &                     &                     &                     &     (0.000)         &                     &                                      \\
[1em]
Distance to Psych&                     &                     &                     &                     &      0.0000         &                                    \\
            &                     &                     &                     &                     &     (0.000)         &                                     \\
[1em]
Km to any border&                     &                     &                     &                     &                     &      -0.0000\sym{**}                        \\
            &                     &                     &                     &                     &                     &     (0.000)                            \\
[1em]
Constant      &     0.0517\sym{***}        &     0.0488\sym{***} &      0.0519\sym{***}        &     0.0476\sym{***}         &    0.0515\sym{***}         &     -0.0529\sym{***}         \\
             &     (0.002)         &     (0.002)         &     (0.002)         &     (0.002)         &     (0.002)         &     (0.000)                 \\
[1em]
\hline
County FE     &         Yes         &         Yes         &         Yes         &         Yes         &         Yes         &         Yes            \\
\hline
$R^2$       &      0.022        &      0.021         &      0.022         &      0.022        &      0.022         &      0.022              \\
\(N\)       &      265719         &      265719         &      263894         &      265719         &      265719         &      265719               \\
\hline\hline
\multicolumn{4}{l}{\textit{Note:} Standard errors in parentheses. $^{*}$p$<$0.1; $^{**}$p$<$0.05; $^{***}$p$<$0.01} \\ 
\end{tabular}}
    \label{tab:my_label}
\end{table}
}

{
\def\sym#1{\ifmmode^{#1}\else\(^{#1}\)\fi}
\begin{table}
    \centering
\caption{Robustness checks of Probability of Antidepressant Use estimation. Logistic regressions.}\label{tab_app_robust2}
\resizebox{\textwidth}{!}{%
\begin{tabular}{l*{7}{c}}
\hline\hline
            &\multicolumn{1}{c}{Model 1}&\multicolumn{1}{c}{Model 2}&\multicolumn{1}{c}{Model 3}&\multicolumn{1}{c}{Model 4}&\multicolumn{1}{c}{Model 5}&\multicolumn{1}{c}{Model 6}\\
\hline
            &\multicolumn{1}{c}{Baseline}&\multicolumn{1}{c}{Strict Definition (ICD-10)}&\multicolumn{1}{c}{d$\geq 10$}&\multicolumn{1}{c}{Friend}&\multicolumn{1}{c}{Psychiatry}&\multicolumn{1}{c}{Borders}\\
\hline
        &                     &                     &                     &                     &                     &                                 \\
$SD_{i}$      &     -0.0291\sym{**}&     -0.0353\sym{**} &     -0.0298\sym{**}&     -0.0256\sym{*}&     -0.0293\sym{**}&     -0.0308\sym{**} \\
            &     (0.014)         &     (0.014)         &     (0.014)         &     (0.014)         &     (0.14)         &     (0.014)     \\
[1em]
$LC_{i}$ &     -0.0117 &     -0.0095         &     -0.0252\sym{*}&     -0.0000 &     -0.0116 &     -0.0123 \\
            &     (0.012)         &     (0.012)         &     (0.014)         &     (0.012)         &     (0.012)         &     (0.012)      \\
[1em]
ln $\left (d_{i} \right )$    &     -0.0432\sym{***}&     -0.0503\sym{***}&     -0.0527\sym{***}&     -0.1068\sym{***}&     -0.0501\sym{***}&     -0.0508\sym{***} \\
            &     (0.013)     &     (0.014)         &     (0.014)         &     (0.016)         &     (0.015)         &     (0.013)       \\
[1em]
$F_{i}^{50}$          &      0.0044  &      0.0160         &      0.0022         &      0.0024         &      0.0047  &      0.0026    \\
            &     (0.014)         &     (0.014)         &     (0.014)         &     (0.014)         &     (0.014)         &     (0.014)     \\
[1em]
Male          &     -0.6278\sym{***}&     -0.6884\sym{***}&     -0.6253\sym{***}&     -0.6200\sym{***}&     -0.6278\sym{***}&     -0.6279\sym{*** }\\
            &     (0.023)         &     (0.000)         &     (0.023)         &     (0.023)         &     (0.023)         &     (0.023)           \\
[1em]
Age         &      0.6875\sym{***}&      0.6803\sym{***}&      0.6880\sym{***}&      0.6711\sym{***}&      0.6876\sym{***}&      0.6872\sym{***} \\
            &     (0.010)         &     (0.000)         &     (0.010)         &     (0.010)         &     (0.010)         &     (0.010)      \\
[1em]
Unemployment$_h$   &      0.0553\sym{**}  &      0.0019\sym{**} &      0.0477\sym{**}  &      0.0480\sym{**}  &      0.0468\sym{**}  &      0.0442\sym{**}  \\
            &     (0.022)         &     (0.023)         &     (0.022)         &     (0.022)         &     (0.022)         &     (0.022)          \\
[1em]
Income$_h$ &     -0.0494\sym{**}&     -0.0367\sym{*}  &     -0.0463\sym{**}&     -0.0398\sym{**}&     -0.04845\sym{**}&     -0.0530\sym{***} \\
            &     (0.019)         &     (0.020)         &     (0.019)         &     (0.019)         &     (0.019)         &     (0.019)             \\
[1em]
Friend takes Antidep.       &                     &                     &                     &      0.2637\sym{***}&                     &           \\
            &                     &                     &                     &     (0.029)         &                     &                   \\
[1em]
Distance to Psych&                     &                     &                     &                     &      0.0003         &             \\
            &                     &                     &                     &                     &     (0.000)         &                   \\
[1em]
Km to any border&                     &                     &                     &                     &                     &      -0.0006\sym{*}                \\
            &                     &                     &                     &                     &                     &     (0.000)                \\
[1em]
Constant      &     -3.1216\sym{***}    &   -3.1872\sym{***} &      -3.1192\sym{***}      &   -3.3019\sym{***}      &   -3.1269\sym{***}     &   -3.1013\sym{***}   \\
            &     (0.056)         &     (0.058)         &     (0.056)         &     (0.060)         &     (0.057)         &     (0.057)           \\
[1em]
\hline
County FE     &         Yes         &         Yes         &         Yes         &         Yes         &         Yes         &         Yes        \\
\hline
\(N\)       &      265719         &      265719         &      263894         &      265719         &      265719         &      265719          \\
\hline\hline
\multicolumn{4}{l}{\textit{Note:} Standard errors in parentheses. $^{*}$p$<$0.1; $^{**}$p$<$0.05; $^{***}$p$<$0.01} \\ 
\end{tabular}}
    \label{tab:my_label}
\end{table}
}

In order to ensure the robustness of our LMP and logistic regression models, we conducted several additional multivariate analyses. Table \ref{tab_app_robust1} with Linear Probability and \ref{tab_app_robust2} with Logit models document these and contain the main model too (Model 1 in both tables). 

In Model 2 of both Tables S5 and Table S6, the indicator of antidepressant use is restricted to those prescriptions on which the first digit of the ICD-10 (International Classification of Diseases) code is ``F'', indicating mental, behavioral and neurodevelopmental disorders. This can be considered as a clean indicator of antidepressant use that is surely related to mental health problems. 
However, the reliability of diagnosis codes on the prescriptions are limited by the fact that diagnosis code on the prescription has no impact on the cost or any other aspects of the medication or on the following therapy. Moreover, a prescription can be used even if the diagnosis code is missing, which is indeed the case for 5\% of the prescriptions in our raw data. In this specification, the negative association between Spatial Diversity and Antidepressant Use are still significant in both the LPM and logistic regression models.

Recognizing the potential bias associated with smaller network sizes, we restricted our sample to individuals who had at least 10 acquaintances in the online social network. The results of this restricted sample analysis are presented in Model 3 in both tables and provide insights into the impact of network size on our findings, we see that all variables remained unchanged. 

Building upon the notion that depression can be influenced by social norms and the behavior of acquaintances, similar to previous research on the influence of friends' obesity on individual mental health \cite{rosenquist2011social}, we incorporated a control for the presence of depressed individuals within our social network. In Model 4, we observed a significant peer effect, indicating the influence of peers on an individual's likelihood of taking antidepressants. Our main variables remain unchanged. However, since our study design lacks temporal information about the friendship ties, establishing causality remains a challenge. This presents an important avenue for future research to explore the dynamics of depression transmission.

In Model 5 of both tables, we examined the role of accessibility to psychiatric clinics, specifically focusing on the distance an individual has to travel to reach the nearest clinic with psychiatric services. Our expectation was that individuals living closer to clinics would have better access to diagnosis and antidepressant medication \cite{tadmon2023differential}. Surprisingly, the results did not confirm this hypothesis, suggesting that proximity to a clinic did not significantly influence the likelihood of depression diagnosis in the presence of social network variables. 

Additionally, we introduced a variable capturing the distance to the nearest border in Model 6 of both tables. Given Hungary's unique historical context, where settlements with Hungarian citizens lie beyond the borders of neighboring countries and fall outside the coverage of the Hungarian health insurance system, we anticipated potential limitations in observing social contacts and network diversity in these border areas. However, the results did not reveal any substantial border effect on the spread of depression.


By conducting these robustness checks, we have further examined the stability and generalizability of our main results reported in the main text and in Supporting Information 5. Despite variations in sample restrictions, the inclusion of additional variables, and the choice of alternative regression methods, our findings consistently support the previously unknown relationship between Spatial Diversity as a measure of bridging social capital and Antidepressant Use, providing confidence in the reliability of our results.
 



\newpage
\section*{Supporting Information 8: Interaction of Local Cohesion and Spatial Diversity}

Interaction terms between variables are commonly used for estimating a variable's conditional effect on another's contribution to the outcome. However, interpreting an interaction term is not as simple as interpreting a coefficient. In the following model, we utilize an interaction term to estimate the conditional effect of a given level of Spatial Diversity $SD_{i}$ on the contribution of Local Cohesion $LC_{i}$ to the prediction of our dependent variable, the probability of antidepressant use:

\begin{equation}
P(A_{i}=1) = \alpha 
+ \beta_{1} LC_{i}
+ \beta_{2} {SD_{i}}
+ \beta_{3} \big ( LC_{i}\times{SD_{i}}   \big )
+ \beta_{4} ln \left (d_{i} \right )
+ \beta_{5} F_{i}^{r}
+ \beta \mathbf{X}_{i}
+ \beta \mathbf{S}_{h},
\end{equation}
where $LC_{i}\times{SD_{i}}$ is the interaction term.

Table \ref{tab_app_interact} presents regression results including interaction terms and contains three models that differ by the way how Local Cohesion variable is normalized. Model 1 contains the Local Cohesion variable that we discuss in the main text and in Figure \ref{figa3}. In Model 2, we use an alternative measure that is the local clustering normalized by degree. Model 3 contains the non-normalized value of local clustering. The negative sign and significance of main explanatory variables are persistent across the models. However, the interaction term is significant only in Model 1 and Model 2. The positive value indicates that Spatial Diversity mitigates the relationship between Local Cohesion and Antidepressant Use.

 \newpage

\begin{table}[!htbp] \centering 
\caption{Probability of Antidepressant Use with interaction terms. Linear probability regression. }\label{tab_app_interact}
\begin{tabular}{@{\extracolsep{5pt}}lccc} 
\\[-1.8ex]\hline 
\hline\hline
            &\multicolumn{1}{c}{Model 1}&\multicolumn{1}{c}{Model 2}&\multicolumn{1}{c}{Model 3}\\
\hline
            &\multicolumn{1}{c}{Erdős-Rényi normalized LC}&\multicolumn{1}{c}{Degree normalized LC}&\multicolumn{1}{c}{Non-normalized LC}\\
\hline \\[-1.8ex] 
 $SD_i$ & $-$0.004$^{***}$ & $-$0.003$^{***}$ & $-$0.003$^{***}$ \\ 
  & (0.000) & (0.000) & (0.000) \\ 
  & & & \\ 
 $LC_i$ & $-$0.002$^{***}$ & $-$0.002$^{***}$ &  $-$0.001$^{**}$\\ 
  & (0.000) &  (0.000)& (0.000) \\ 
  & & & \\ 
 $LC_{i_{k}}\times{SD_{i}}$ & 0.001$^{***}$ & 0.001$^{**}$ & 0.000 \\ 
  & (0.000) & (0.000) & (0.000) \\ 
  & & & \\ 
 ln $\left (d_{i} \right )$ & $-$0.006$^{***}$ & $-$0.005$^{***}$ & $-$0.005$^{***}$ \\ 
  & (0.000) & (0.000) & (0.000) \\ 
  & & & \\ 
 $F_{i}^{50}$ & $-$0.001 & $-$0.000 & $-$0.000 \\ 
  & (0.000) & (0.000) & (0.000) \\ 
  & & & \\ 
 Male$_i$ & $-$0.020$^{***}$ & $-$0.020$^{***}$ & $-$0.020$^{***}$ \\ 
  & (0.001) & (0.001) & (0.001) \\ 
  & & & \\ 
 Age$_i$ & 0.026$^{***}$ & 0.026$^{***}$ & 0.026$^{***}$ \\ 
  & (0.000) & (0.000) & (0.000) \\ 
  & & & \\ 
 Income$_h$ & $-$0.001$^{**}$ & $-$0.002$^{**}$ & $-$0.002$^{***}$ \\ 
  & (0.001) & (0.001) & (0.001) \\ 
  & & & \\ 
 Unemployment$_h$ & 0.002$^{**}$ & 0.002$^{**}$ & 0.002$^{**}$ \\ 
  & (0.001) & (0.001) & (0.001) \\ 
  & & & \\ 
 Constant & 0.053$^{***}$ & 0.053$^{***}$ & 0.053$^{***}$ \\ 
  & (0.003) & (0.003) & (0.003) \\ 
  & & & \\ 
Observations & 265,719 & 272,480 & 272,480 \\ 
R$^{2}$ & 0.022 & 0.022 & 0.022 \\ 
Adjusted R$^{2}$ & 0.022 & 0.022 & 0.022 \\ 
\hline\hline
\multicolumn{4}{l}{\textit{Note:} Standard errors in parentheses. $^{*}$p$<$0.1; $^{**}$p$<$0.05; $^{***}$p$<$0.01} \\ 
\end{tabular} 
\end{table}

However, we are not  directly interested in the coefficients and the standard errors of the model parameters $\beta_{1}$, $\beta_{2}$ and $\beta_{3}$ etc., per se. In this case the contribution  of $LC_{i}$ to $P_{i}$ could be estimated by the marginal effect:

\begin{equation}
    \frac{\delta P_{i}}{\delta  LC_{i}}=\beta_{1}+\beta_{3}SD_{i},
\end{equation}
while the standard error of marginal effect is estimated as:
\begin{equation}
    \hat{\sigma}\bigg(\frac{\delta P_{i}}{\delta LC_{i}}\bigg)=\sqrt{var(\hat{\beta_{1}})
    + SD_{i}^{'2}(\hat{\beta_{3}})
    + 2SD_{i}cov(\hat{\beta_{1}},\hat{\beta_{3}})
    }.
\end{equation}
Based on Equation S4, previous statistical studies 
have shown that we cannot rule out the possibility of a statistically significant contribution of $LC_{i}$ on $P_{i}$ for certain values of $SD_{i}$, even if all other model parameters are insignificant. This means that we cannot determine the real conditional effect of $LC_{i}$ on $P_{i}$ solely based on the effect sizes and standard errors of $\beta_{1}$ and $\beta_{3}$. In order to address this issue, we follow the literature 
and calculate the marginal effect of $LC_{i}$ at all possible values of $SD_{i}$.

\begin{figure*}[!ht]
\centering\includegraphics[width=\textwidth]{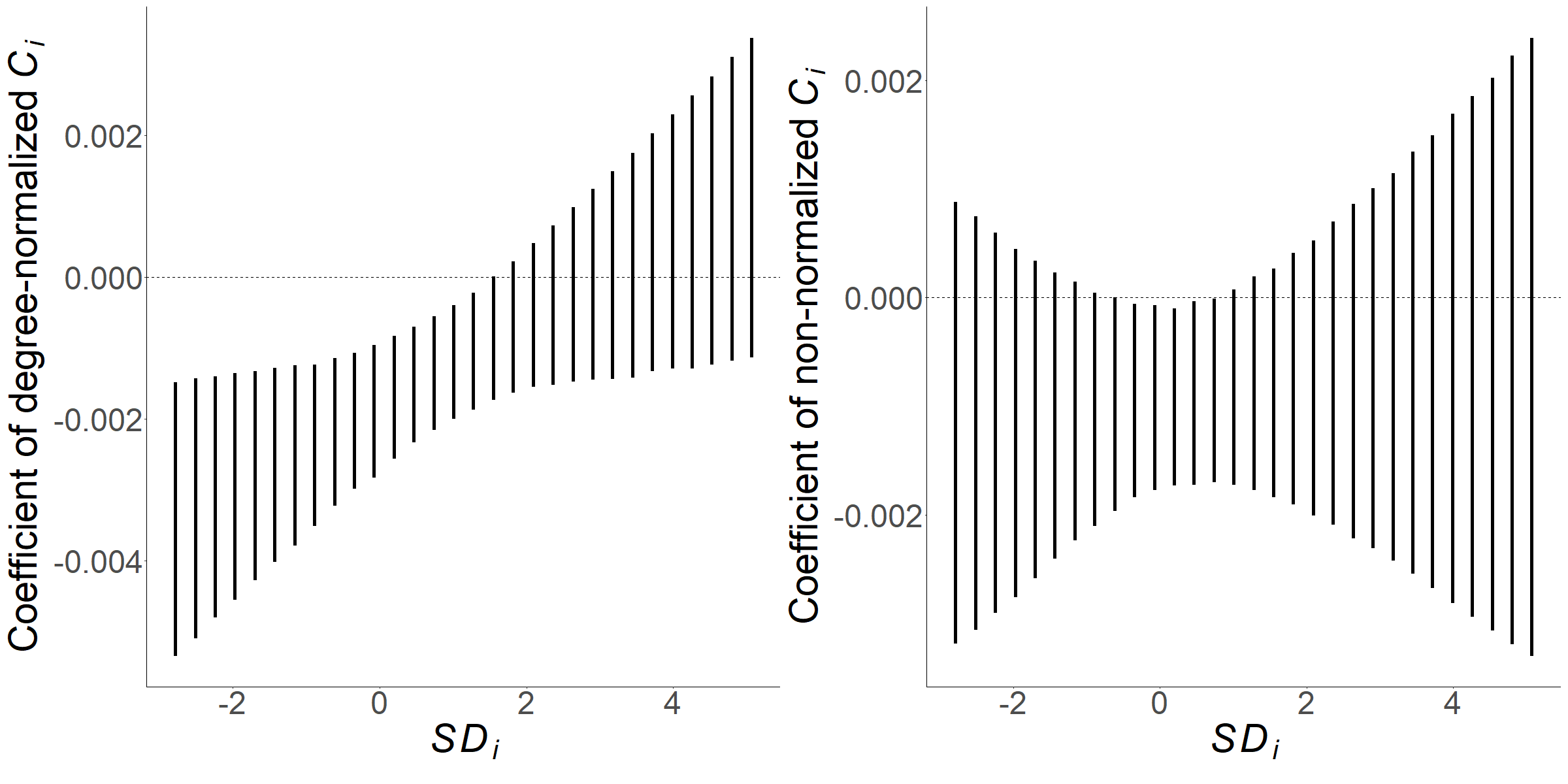}
\caption{Marginal effect of alternative measures of Local Cohesion on the Probability of Antidepressant Use by levels of Spatial Diversity.}
\label{figa6}
\end{figure*}

The main text contains the marginal effect of Local Cohesion on Antidepressant Use at levels of Spatial Diversity. Figure S7 of the Appendix complements this with Local Cohesion indices normalized or not. The degree-normalized $LC_{i}$ indicator behaves similarly as the Erdős-Rényi normalized variable. This result implies that the importance of Local Cohesion decreases and becomes insignificant as the value of Spatial Diversity increases. However, the non-normalized version of local clustering does not show this pattern.

\newpage
\section*{Supporting Information 9: Access to Antidepressants as Selection Mechanism}

The accessibility to healthcare might introduce a selection bias because those who have higher accessibility are more likely to purchase antidepressants; consequently, antidepressant use might be systematically under-measured in low accessibility communities. To control for this selection bias, we measure the distance to the nearest psychiatric center, that has been found to increase medical treatment intensity \cite{mcclellan1994does, tadmon2023differential}. We argue that the ratio of antidepressant users in the region of the town is also a good measure to sort out the accessibility bias, for two reasons. First, only small-town residents are included in the analysis; thus, they do not have a dominant weight among antidepressant users in their larger regions. Second, the ratio of antidepressant users is spatially correlated such that high versus low ratio areas are far from each other, as it was reported in Figure 1B. 

To mitigate the selection bias of unequal access to antidepressants, we estimate the probability of antidepressant use $A_i$ with the following logistic regression

\begin{equation}
    ln(\frac{P(A_i=1)}{1-P(A_i=1)})=\beta_{1}+\beta_{2}Dist_{h}+\beta_{3}A_{r},
\end{equation}
where $Dist_{h}$ is the Euclidean distance from home-town $h$ of patient $i$ to the closest psychiatric centre and $A_{r}$ is the ratio of antidepressant users in the home-region $r$ of patient $i$ ($h \in r$).

Table S8 demonstrates that, unlike in previous research where distance to healthcare institutions were correlated negatively with treatment intensity, $Dist_{h}$ has a significant positive, but weak correlation with the probability of antidepressant use. The reason behind this finding is that psychiatric centers are located in regional centres and antidepressant use is typically higher in more remote towns that are further away from these central places. Next, we find that $A_{r}$ is a significant and strong predictor of antidepressant use.

\begin{table}[!htbp] \centering 
  \caption{Antidepressant use, distance to psychiatric centers and usage rate in the region} 
  \label{} 
\begin{tabular}{@{\extracolsep{5pt}}lccc} 
\\[-1.8ex]\hline 
\hline \\[-1.8ex] 
 & \multicolumn{3}{c}{\textit{Dependent variable:}} \\ 
\cline{2-4} 
\\[-1.8ex] & \multicolumn{3}{c}{Probability of Antidepressant Use} \\ 
 \\[-1.8ex] & (1) & (2) & (3)\\ 
\hline \\[-1.8ex] 
 Distance to psychiatric centers (km) & 0.0005 &  & 0.0002 \\ 
  & (0.001) &  & (0.001) \\ 
  & & & \\ 
 Antidepressant usage rate in the region (log) &  & 0.892$^{***}$ & 0.891$^{***}$ \\ 
  &  & (0.054) & (0.054) \\ 
  & & & \\ 
 Constant & $-$3.272$^{***}$ & $-$0.417$^{**}$ & $-$0.426$^{**}$ \\ 
  & (0.022) & (0.173) & (0.175) \\ 
  & & & \\ 
\hline \\[-1.8ex] 
Observations & 277,344 & 277,344 & 277,344 \\ 
\hline 
\hline \\[-1.8ex] 
\multicolumn{4}{l}{\textit{Note:} Standard errors in parentheses. $^{*}$p$<$0.1; $^{**}$p$<$0.05; $^{***}$p$<$0.01} \\ 
\end{tabular} 
\end{table}

\begin{figure*}[!h]
\centering\includegraphics[width=0.8\textwidth]{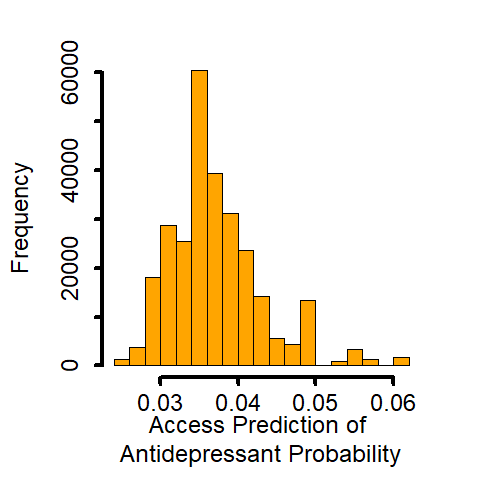}
\caption{Distribution of predicted values of antidepressant use by Distance to psychiatric centers and Antidepressant usage rate in the region of the individual.}
\label{figa9}
\end{figure*}

Next, we predict the probability of antidepressant use by using the point-estimates of $Dist_h$ and $A_r$. Figure S8 illustrates the distribution of the predicted value ($\hat{A_h}$). By including $\hat{A_h}$ in the probability of $A_i$ estimations in the main text, we control for the potential bias of the spatial biases of antidepressant access. In Table S9, we confirm that this inclusion does not eliminate the sign and significance of our predictors that we reported in Figure 3 in the main text. It however does decrease the significance of the predictors when regressing the Days of Treatment. Therefore, we include $\hat{A_h}$ in the dynamics of dosage regressions that we report in Table 1 in the main text.

\begin{table}[!htbp] \centering 
  \caption{Inclusion of predicted probability of antidepressant use as control variable} 
  \label{} 
\begin{tabular}{@{\extracolsep{5pt}}lccc} 
\\[-1.8ex]\hline 
\hline \\[-1.8ex] 
\\[-1.8ex] & Probability of Antidepressant & \multicolumn{2}{c}{Days of Therapy in 2011} \\ 
 &  & All obs & Antidepressant Users  \\ 
\\[-1.8ex] & (1) & (2) & (3)\\ 
\hline \\[-1.8ex] 
 $SD_i$ & $-$0.003$^{***}$ & $-$0.018$^{***}$ & $-$0.026$^{*}$ \\ 
  & (0.000) & (0.002) & (0.015) \\ 
  & & & \\ 
 $LC_i$ & $-$0.001$^{**}$ & $-$0.006$^{**}$ & $-$0.000 \\ 
  & (0.000) & (0.002) & (0.013) \\ 
  & & & \\ 
 ln($d_i$) & $-$0.005$^{***}$ & $-$0.029$^{***}$ & $-$0.016 \\ 
  & (0.000) & (0.002) & (0.014) \\ 
  & & & \\ 
 $F^50$ & $-$0.000 & $-$0.003 & $-$0.016 \\ 
  & (0.000) & (0.002) & (0.014) \\ 
  & & & \\ 
 Male$_i$ & $-$0.020$^{***}$ & $-$0.103$^{***}$ & $-$0.095$^{***}$ \\ 
  & (0.001) & (0.004) & (0.024) \\ 
  & & & \\ 
 Age$_i$ & 0.026$^{***}$ & 0.133$^{***}$ & 0.168$^{***}$ \\ 
  & (0.000) & (0.002) & (0.013) \\ 
  & & & \\ 
 Income$_h$ & $-$0.001$^{**}$ & $-$0.007$^{**}$ & $-$0.017 \\ 
  & (0.001) & (0.003) & (0.020) \\ 
  & & & \\ 
 Unemployment$_h$ & 0.001 & 0.004 & $-$0.013 \\ 
  & (0.001) & (0.004) & (0.023) \\ 
  & & & \\ 
 $\hat{A_h}$ & 1.100$^{***}$ & 5.539$^{***}$ & 1.939 \\ 
  & (0.091) & (0.456) & (2.436) \\ 
  & & & \\ 
 Constant & 0.011$^{**}$ & 0.053$^{**}$ & 4.852$^{***}$ \\ 
  & (0.004) & (0.022) & (0.116) \\ 
  & & & \\ 
Observations & 265,719 & 265,719 & 9,769 \\ 
R$^{2}$ & 0.023 & 0.024 & 0.028 \\ 
\hline \\[-1.8ex] 
\multicolumn{4}{l}{\textit{Note:} Standard errors in parentheses. $^{*}$p$<$0.1; $^{**}$p$<$0.05; $^{***}$p$<$0.01}  \\ 
\end{tabular} 
\end{table} 

\newpage
\section*{Supporting Information 10: Dynamics of Days of Therapy}

We observe the quantity of purchased antidepressants that enables us to examine the impact of Spatial Social Capital on the dynamics of antidepressant use by focusing on antidepressant users only. $Z_{i,t}$ is the Days of Therapy (DOT). To obtain the DOT indicator, for each antidepressant type we multiply the volume of  antidepressant packages purchased in year $t$ by individual $i$ with the per-package DOT value (as included in our data), and add up these products. The distribution of the log-transformed $Z_{i,t}$ illustrates that most patients purchase a quantity that covers almost the entire year (the median is around $10^{2.5} = 315$ days).

\begin{figure*}[!ht]
\centering\includegraphics[width=0.6\textwidth]{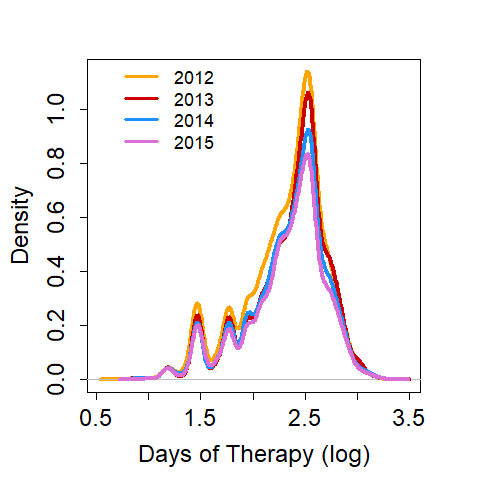}
\caption{Distribution of Days of Therapy reveals similar patterns over the analyzed period.}
\label{figa8}
\end{figure*}
$Z_{i,t}$ is strongly correlated across subsequent years (reported in Figure 2D in the main text). Thus, running regressions on the level of $Z_{i,t}$ by controlling for the value of $Z_{i,2011}$ enables us to evaluate the forms of Spatial Social Capital in mitigating the dose of antidepressants. We test the following equation with OLS regressions for $t \in \big\{2012, 2013, 2014, 2015\}$:

\begin{equation}
Z_{i,t} = \alpha 
+ \beta_{1} Z_{i,2011}
+ \beta_{2} LC_{i}
+ \beta_{3} {SD_{i}}
+ \beta_{4} ln \left (d_{i} \right )
+ \beta_{5} F_{i}^{r}
+ \beta \mathbf{X}_{i}
+ \beta \mathbf{S}_{h}
+ \mathbf{D}_{c}
+ \varepsilon_{i,t},
\end{equation}
where  $Z_{i,2011}$ is the Days of Therapy in year 2011, the other co-variates are identical to the ones that we used in preceding analyses, and $\varepsilon_{i,t}$ is the error term.

As expected, $Z_{i,2011}$ is strongly correlated with $Z_{i,t}$ for every $t$ (Table \ref{tab_app_dynamics}). We also find the expected significant negative coefficient of the Male variable and the significant positive coefficient of the Age variable. Degree has a negative coefficient but is only significant at the 5\% level only at $t=2014$.

We find that Local Cohesion does not but Spatial Diversity does have a significant negative correlation with $Z_{i,t}$. This finding confirm that bridging social capital has a mitigation effect on the mental disorders that is reflected by decreasing Days of Therapy.

Next, we regress $\Delta Z_{i,t}$ that is the change of Days of Therapy between year 2011 and $t$, using the same variables: 

\begin{equation}
\Delta Z_{i,t} = \alpha 
+ \beta_{1} Z_{i,2011}
+ \beta_{2} LC_{i}
+ \beta_{3} {SD_{i}}
+ \beta_{4} ln \left (d_{i} \right )
+ \beta_{5} F_{i}^{r}
+ \beta \mathbf{X}_{i}
+ \beta \mathbf{S}_{k}
+ \mathbf{D}_{c}
+ \varepsilon_{i,t}.
\end{equation}

These final results confirm previous findings (Table \ref{tab_app_delta}). The correlation of $Z_{i,2011}$ drops as well as the $R^2$ of the models. Yet, the coefficients of Spatial Diversity, Degree, Male, and Age remain almost unchanged.

\begin{table}[!htbp] 
\centering 
  \caption{Days of Therapy (DOT) in subsequent years. Linear regression controlling for DOT in 2011.} 
  \label{tab_app_dynamics} 
\begin{tabular}{@{\extracolsep{5pt}}lcccc} 
\\[-1.8ex]\hline 
\hline \\[-1.8ex] 

 & 2012 & 2013 & 2014 & 2015 \\ 
\\[-1.8ex] & (1) & (2) & (3) & (4)\\ 
\hline \\[-1.8ex] 
 Antidepressant DOT 2011 & 1.498$^{***}$ & 1.393$^{***}$ & 1.300$^{***}$ & 1.222$^{***}$ \\ 
  & (0.018) & (0.020) & (0.021) & (0.022) \\ 
  & & & & \\ 
 Spatial Diversity & $-$0.054$^{**}$ & $-$0.071$^{**}$ & $-$0.075$^{**}$ & $-$0.071$^{**}$ \\ 
  & (0.027) & (0.031) & (0.033) & (0.035) \\ 
  & & & & \\ 
 Local Cohesion & $-$0.009 & $-$0.023 & $-$0.012 & $-$0.019 \\ 
  & (0.024) & (0.027) & (0.031) & (0.032) \\ 
  & & & & \\ 
 Degree & $-$0.013 & $-$0.057$^{*}$ & $-$0.063$^{**}$ & $-$0.043 \\ 
  & (0.026) & (0.030) & (0.031) & (0.033) \\ 
  & & & & \\ 
 Friends within 50 km & 0.032 & $-$0.014 & $-$0.020 & $-$0.036 \\ 
  & (0.026) & (0.030) & (0.031) & (0.032) \\ 
  & & & & \\ 
 Male & $-$0.210$^{***}$ & $-$0.247$^{***}$ & $-$0.316$^{***}$ & $-$0.302$^{***}$ \\ 
  & (0.044) & (0.050) & (0.052) & (0.054) \\ 
  & & & & \\ 
 Age & 0.217$^{***}$ & 0.334$^{***}$ & 0.359$^{***}$ & 0.312$^{***}$ \\ 
  & (0.023) & (0.027) & (0.028) & (0.029) \\ 
  & & & & \\ 
 Income per capita & $-$0.023 & 0.025 & 0.062 & $-$0.002 \\ 
  & (0.037) & (0.043) & (0.044) & (0.045) \\ 
  & & & & \\ 
 Unemployment rate & $-$0.010 & 0.021 & 0.058 & $-$0.015 \\ 
  & (0.043) & (0.049) & (0.051) & (0.053) \\ 
  & & & & \\ 
  & & & & \\ 
 Constant & $-$4.152$^{***}$ & $-$4.064$^{***}$ & $-$3.618$^{***}$ & $-$3.250$^{***}$ \\ 
  & (0.153) & (0.176) & (0.182) & (0.188) \\ 
  & & & & \\ 
Observations & 9,769 & 9,769 & 9,769 & 9,769 \\ 
R$^{2}$ & 0.447 & 0.355 & 0.317 & 0.276 \\ 
Adjusted R$^{2}$ & 0.445 & 0.353 & 0.315 & 0.274 \\ 
\hline \\[-1.8ex] 
\multicolumn{4}{l}{\textit{Note:} Standard errors in parentheses. $^{*}$p$<$0.1; $^{**}$p$<$0.05; $^{***}$p$<$0.01} & \\
\end{tabular} 
\end{table} 

\begin{table}[!htbp] \centering 
  \caption{Delta Days of Therapy (DOT, $Z_{i,t}$) in subsequent years. Linear regression controlling for DOT in 2011.} 
  \label{tab_app_delta} 
\begin{tabular}{@{\extracolsep{5pt}}lcccc} 
\\[-1.8ex]\hline 
\hline \\[-1.8ex] 
\\[-1.8ex] & $\Delta Z_{i,2012}$ & $\Delta Z_{i,2013}$ & $\Delta Z_{i,2014}$ & $\Delta Z_{i,2015}$ \\
\\[-1.8ex] & (1) & (2) & (3) & (4)\\ 
\hline \\[-1.8ex] 
 Antidepressant DOT 2011 & 0.498$^{***}$ & 0.393$^{***}$ & 0.300$^{***}$ & 0.222$^{***}$ \\ 
  & (0.018) & (0.020) & (0.021) & (0.022) \\ 
  & & & & \\ 
 Spatial Diversity & $-$0.054$^{**}$ & $-$0.071$^{**}$ & $-$0.071$^{**}$ & $-$0.070$^{**}$ \\ 
  & (0.027) & (0.031) & (0.032) & (0.033) \\ 
  & & & & \\ 
 Local Cohesion & $-$0.009 & $-$0.023 & $-$0.005 & $-$0.019 \\ 
  & (0.024) & (0.027) & (0.028) & (0.029) \\ 
  & & & & \\ 
 Degree & $-$0.013 & $-$0.057$^{*}$ & $-$0.061$^{**}$ & $-$0.043 \\ 
  & (0.026) & (0.030) & (0.031) & (0.032) \\ 
  & & & & \\ 
 Friends within 50 km & 0.032 & $-$0.014 & $-$0.019 & $-$0.036 \\ 
  & (0.026) & (0.030) & (0.031) & (0.032) \\ 
  & & & & \\ 
 Male & $-$0.210$^{***}$ & $-$0.247$^{***}$ & $-$0.317$^{***}$ & $-$0.302$^{***}$ \\ 
  & (0.044) & (0.050) & (0.052) & (0.053) \\ 
  & & & & \\ 
 Age & 0.217$^{***}$ & 0.334$^{***}$ & 0.359$^{***}$ & 0.312$^{***}$ \\ 
  & (0.023) & (0.027) & (0.028) & (0.029) \\ 
  & & & & \\ 
 Income per capita & $-$0.023 & 0.025 & 0.061 & $-$0.002 \\ 
  & (0.037) & (0.043) & (0.044) & (0.045) \\ 
  & & & & \\ 
 Unemployment rate & $-$0.010 & 0.021 & 0.058 & $-$0.015 \\ 
  & (0.043) & (0.049) & (0.051) & (0.053) \\ 
  & & & & \\ 
 Constant & $-$4.152$^{***}$ & $-$4.064$^{***}$ & $-$3.616$^{***}$ & $-$3.250$^{***}$ \\ 
  & (0.153) & (0.176) & (0.182) & (0.188) \\ 
  & & & & \\ 
Observations & 9,769 & 9,769 & 9,769 & 9,769 \\ 
R$^{2}$ & 0.102 & 0.069 & 0.055 & 0.036 \\ 
Adjusted R$^{2}$ & 0.100 & 0.066 & 0.052 & 0.033 \\ 
\hline \\[-1.8ex] 
\multicolumn{4}{l}{\textit{Note:} Standard errors in parentheses. $^{*}$p$<$0.1; $^{**}$p$<$0.05; $^{***}$p$<$0.01} & \\
\end{tabular} 
\end{table} 

\begin{table}[!htbp] \centering 
  \caption{Delta Days of Therapy (DOT, $Z_{i,t}$) in subsequent years. Linear regression controlling for DOT in 2011 and for Antidepressant Access.} 
  \label{} 
\begin{tabular}{@{\extracolsep{5pt}}lcccc} 
\\[-1.8ex]\hline 
\hline \\[-1.8ex] 
\\[-1.8ex] & DOT 2012 & DOT 2013 & DOT 2014 & DOT 2015\\ 
\hline \\[-1.8ex] 
 DOT 2011 & 1.497$^{***}$ & 1.392$^{***}$ & 1.299$^{***}$ & 1.221$^{***}$ \\ 
  & (0.018) & (0.020) & (0.021) & (0.022) \\ 
 Antidepressant Access & 4.829 & 9.443$^{*}$ & 11.284$^{**}$ & 15.183$^{***}$ \\ 
  & (4.482) & (5.165) & (5.325) & (5.505) \\ 
\hline \\[-1.8ex] 
 $SD_i$ & $-$0.104$^{***}$ & $-$0.125$^{***}$ & $-$0.147$^{***}$ & $-$0.140$^{***}$ \\ 
  & (0.036) & (0.041) & (0.042) & (0.044) \\ 
 $LC_i$ & $-$0.011 & $-$0.025 & $-$0.008 & $-$0.020 \\ 
  & (0.024) & (0.027) & (0.028) & (0.029) \\ 
 $ln(d_i)$ & $-$0.011 & $-$0.052$^{*}$ & $-$0.057$^{*}$ & $-$0.038 \\ 
  & (0.026) & (0.030) & (0.031) & (0.032) \\ 
 $F_{50}$ & 0.035 & $-$0.009 & $-$0.014 & $-$0.030 \\ 
  & (0.026) & (0.030) & (0.031) & (0.032) \\ 
\hline \\[-1.8ex] 
 Male$_i$ & $-$0.213$^{***}$ & $-$0.249$^{***}$ & $-$0.321$^{***}$ & $-$0.306$^{***}$ \\ 
  & (0.044) & (0.050) & (0.052) & (0.053) \\ 
 Age$_i$ & 0.218$^{***}$ & 0.335$^{***}$ & 0.360$^{***}$ & 0.313$^{***}$ \\ 
  & (0.023) & (0.027) & (0.028) & (0.029) \\ 
\hline \\[-1.8ex] 
 Income$_h$ & $-$0.027 & 0.019 & 0.055 & $-$0.011 \\ 
  & (0.037) & (0.043) & (0.044) & (0.045) \\ 
 Unemployment$_h$ & $-$0.019 & 0.007 & 0.041 & $-$0.038 \\ 
  & (0.043) & (0.050) & (0.051) & (0.053) \\ 
\hline \\[-1.8ex] 
 $SD_i$ × Male$_i$ & 0.058 & 0.010 & 0.093$^{*}$ & 0.085 \\ 
  & (0.046) & (0.053) & (0.055) & (0.056) \\ 
 $SD_i$ × Age$_i$ & 0.048$^{**}$ & 0.074$^{***}$ & 0.073$^{**}$ & 0.067$^{**}$ \\ 
  & (0.024) & (0.028) & (0.029) & (0.030) \\ 
 $SD_i$ × Income$_h$ & $-$0.029 & $-$0.048$^{*}$ & $-$0.027 & $-$0.036 \\ 
  & (0.023) & (0.026) & (0.027) & (0.028) \\ 
 Constant & $-$4.326$^{***}$ & $-$4.410$^{***}$ & $-$4.034$^{***}$ & $-$3.817$^{***}$ \\ 
  & (0.230) & (0.265) & (0.273) & (0.282) \\ 
\hline \\[-1.8ex] 
Observations & 9,769 & 9,769 & 9,769 & 9,769 \\ 
R$^{2}$ & 0.447 & 0.355 & 0.318 & 0.277 \\ 
Adjusted R$^{2}$ & 0.445 & 0.353 & 0.316 & 0.274 \\ 
\hline\hline \\[-1.8ex] 
\multicolumn{4}{l}{\textit{Note:} Standard errors in parentheses. $^{*}$p$<$0.1; $^{**}$p$<$0.05; $^{***}$p$<$0.01} & \\
\end{tabular} 
\end{table} 

\end{document}